\documentclass[]{assets/matterlab}

\usepackage{booktabs}
\usepackage{graphicx}
\usepackage{listings}
\usepackage{amsmath}
\usepackage{tabularx}
\usepackage{listings}
\usepackage{changepage}
\usepackage[inline]{enumitem}

\correspondence{Al\'an Aspuru-Guzik at \email{alan@aspuru.com}}

\metadata[Code]{\url{https://github.com/aspuru-guzik-group/TreeWriter}}

\newcommand{\prose}{prose}

\title{TreeWriter: AI-Assisted Hierarchical Planning and Writing for Long-Form Documents}
\author[1,2]{Zijian Zhang}
\author[1]{Fangshi Du}
\author[1]{Xingjian Liu}
\author[1,2]{Pan Chen}
\author[1]{Oliver Huang}
\author[1]{Runlong Ye}
\author[4]{Michael Liut}
\author[1,2,3,5,6,7,8,9,*]{Al\'an Aspuru-Guzik}
\newcommand{\addressCHEM}{Department of Chemistry, University of Toronto, Lash Miller Chemical Laboratories, 80 St. George Street, ON M5S 3H6, Toronto, Canada}
\newcommand{\addressAC}{Acceleration Consortium, 700 University Ave., M7A 2S4, Toronto, Canada}
\newcommand{\addressCS}{Department of Computer Science, University of Toronto, Sandford Fleming Building, 10 King’s College Road, ON M5S 3G4, Toronto, Canada}
\newcommand{\addressVECTOR}{Vector Institute for Artificial Intelligence, 661 University Ave. Suite 710, ON M5G 1M1, Toronto, Canada}
\newcommand{\addressMSE}{Department of Materials Science \& Engineering, University of Toronto, 184 College St., M5S 3E4, Toronto, Canada}
\newcommand{\addressCHEMENG}{Department of Chemical Engineering \& Applied Chemistry, University of Toronto, 200 College St. ON M5S 3E5, Toronto, Canada}
\newcommand{\addressCIFAR}{Canadian Institute for Advanced Research (CIFAR), 661 University Ave., M5G 1M1, Toronto, Canada}
\newcommand{\addressNVIDIA}{NVIDIA, 431 King St W \#6th, M5V 1K4, Toronto, Canada}

\newcommand{\addressUTM}{Department of Mathematical and Computational Sciences, University of Toronto Mississauga, 3359 Mississauga Road, Deerfield Hall, ON L5L 1C6, Mississauga, Canada}
\renewcommand{\cite}{\citep}
\affiliation[1]{\addressCS}
\affiliation[2]{\addressVECTOR}
\affiliation[3]{\addressCHEM}
\affiliation[4]{\addressUTM}
\affiliation[5]{\addressMSE}
\affiliation[6]{\addressCHEMENG}
\affiliation[7]{\addressAC}
\affiliation[8]{\addressCIFAR}
\affiliation[9]{\addressNVIDIA}

\begin{document}

\renewcommand{\sectionautorefname}{Section}
\renewcommand{\subsectionautorefname}{Section}
\renewcommand{\subsubsectionautorefname}{Section}

\abstract{
Long documents pose many challenges to current intelligent writing systems. These include maintaining consistency across sections, sustaining efficient planning and writing as documents become more complex, and effectively providing and integrating AI assistance to the user. Existing AI co-writing tools offer either inline suggestions or limited structured planning, but rarely support the entire writing process that begins with high-level ideas and ends with polished \prose, in which many layers of planning and outlining are needed. Here, we introduce TreeWriter, a hierarchical writing system that represents documents as trees and integrates contextual AI support. TreeWriter allows authors to create, save, and refine document outlines at multiple levels, facilitating drafting, understanding, and iterative editing of long documents. A built-in AI agent can dynamically load relevant content, navigate the document hierarchy, and provide context-aware editing suggestions. A within-subject study ($N=12$) comparing TreeWriter with Google Docs + Gemini on long-document editing and creative writing tasks shows that TreeWriter improves idea exploration/development, AI helpfulness, and perceived authorial control. A two-month field deployment ($N=8$) further demonstrated that hierarchical organization supports collaborative writing. Our findings highlight the potential of hierarchical, tree-structured editors with integrated AI support and provide design guidelines for future AI-assisted writing tools that balance automation with user agency.
}
\maketitle
\begin{figure}
  \includegraphics[width=\textwidth]{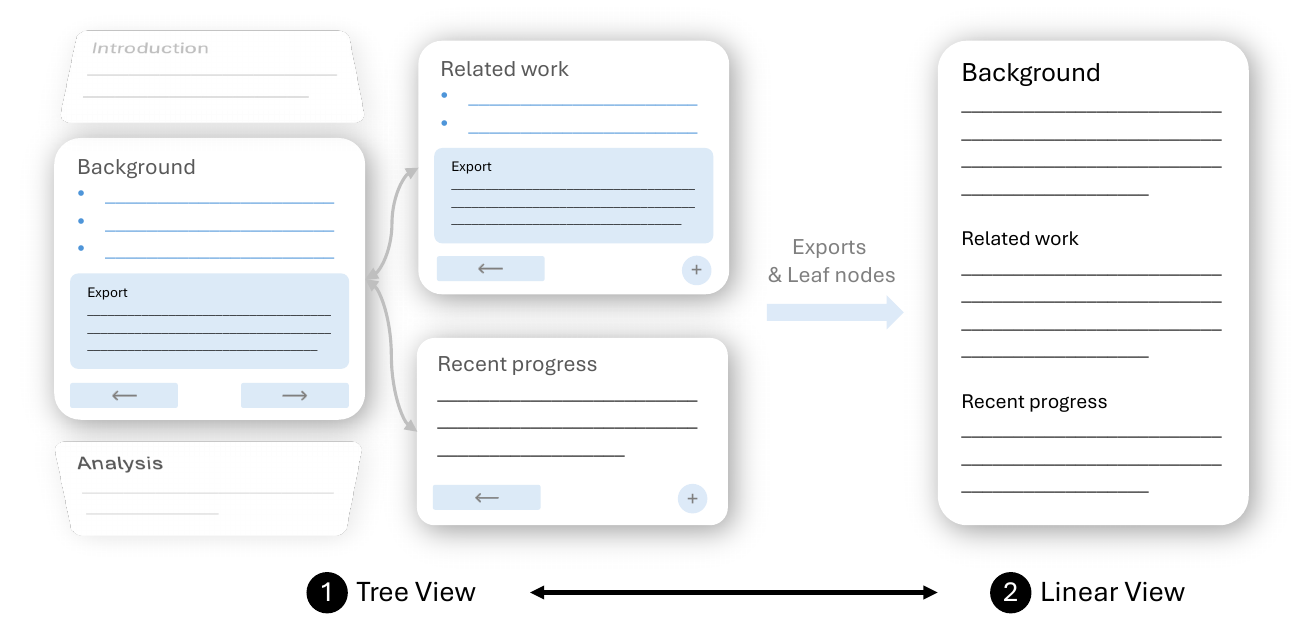}
  \caption{TreeWriter enables users to view and edit their documents in two complementary views: (1) the \textbf{tree view} and (2) the \textbf{linear view}. In the tree view, users develop outlines at each node, expand them into text for the final document, or split them into child nodes for further elaboration. Users can use integrated AI features to maintain coherence and consistency across related nodes. The hierarchical structure formed in this process supports easy navigation and multi-level editing. The linear view compiles complete sections by traversing the subtrees and concatenating the exported content from nodes sequentially. A chat-based writing assistant with a scoped context is available in both views, offering node-level writing and revision suggestions.
  }
\end{figure}

\section{Introduction}

AI-assisted long-form writing presents unique challenges that extend beyond simple text generation, as this task requires managing information at scales that exceed human working memory \cite{Kellogg}. Unlike short-form writing, where authors can maintain a coherent mental representation of the entire text, long documents such as research grants, reports, or books demand sustained attention to structure, coherence, and dependencies across multiple sections. Writers often employ external cognitive scaffolds to mitigate these limitations. Common strategies include repeated readings to construct a cognitive map of the global text structure, adding annotations to the actual texts, and using professional writing tools. In collaborative contexts, the difficulty shifts toward sustaining a shared cognitive representation of the document, which requires aligning understanding across contributors. Collectively, these constraints underscore the need for AI systems that extend beyond local text generation, facilitating the higher-order processes of memory management, which is essential for effective long-form writing. 

The recent advances in large language models (LLMs) create an opportunity to solve these challenges \cite{openai2025introducinggpt4-1, openai2025introducinggpt5, guo2025deepseek, yang2025qwen3, anthropic2025claude4, comanici2025gemini}. 
Text can now be rapidly and automatically reconfigured across multiple representational forms, such as bullet points, continuous prose, or images, to facilitate human understanding and operating \cite{zhang2025treereader}. 
Current AI-assisted writing tools \cite{reza2025co} can be broadly categorized into two paradigms: \textit{inline suggestion editors} that provide immediate text modifications within linear documents~\cite{wordcraft, copoet, abscribe, ghostwriter}, and \textit{conceptual editors} that support planning and organization beyond direct text manipulation~\cite{luminate, visar, siddiqui2025script}. While inline editors offer immediate assistance, they rely heavily on the ability of LLMs to reconstruct the hidden conceptual structure behind texts. Existing conceptual editors, although they expose the conceptual structure to the user, still lack methods to build further hierarchies of abstraction that support the efficient editing of longer documents (See \autoref{sec:related_work}).

To address these limitations, we introduce \textbf{TreeWriter}, a hierarchical writing system that integrates AI assistance with tree-based document organization. TreeWriter represents a document as a hierarchical tree, allowing authors to iteratively construct higher-level abstractions from detailed content with the help of AI, forming a multi-level representation of their document. Based on the tree structure, TreeWriter integrates a writing assistant that can operate on these levels, offering suggestions at multiple levels that maintain consistency between high-level plans and low-level text, thereby supporting both high-level conceptual reasoning and actual document editing throughout the writing process.

The development of the system was informed by a formative study involving six researchers who regularly produce long-form content. The study revealed three key challenges in AI-assisted writing: (1) structuring ideas and transforming them into text, (2) maintaining consistency across sections, and (3) verifying the accuracy and trustworthiness of AI-generated content. These findings informed the design of TreeWriter, focusing on three key objectives: supporting idea structuring and iterative drafting, ensuring cross-section consistency, and facilitating trustworthy AI collaboration.

We evaluated TreeWriter through both a comparative lab study and a two-month field deployment. In the lab study, we compared TreeWriter with Google Docs \cite{google_docs} equipped with the Gemini assistant across two tasks: modifying a long article (approximately 4,000 words) and writing a new article (approximately 800 words). The results demonstrate that TreeWriter effectively supports both editing and drafting of long-form documents while allowing users to retain control over AI-generated content. In the field deployment, participants utilized TreeWriter in real-world collaborative writing projects over two months. Findings from this study suggest that TreeWriter can enhance team collaboration in co-authoring long documents.

The main contributions of this work are:

\begin{adjustwidth}{2em}{}
\begin{enumerate}
    \item The design and implementation of \textbf{TreeWriter}, a novel writing system that demonstrates how a hierarchical document structure can be integrated with AI assistance to overcome the limitations of existing writing tools.
    \item A \textbf{mixed-methods evaluation}, which includes a controlled in-lab user study and a field deployment, demonstrates that TreeWriter is more effective than a standard linear editor for long document editing, creative writing and collaborative writing in many dimensions.
    \item A set of \textbf{design recommendations} for developing AI-assisted writing tools that support long-form document editing and balance automation with user control, informed by insights from our formative study and evaluation findings.
\end{enumerate}
\end{adjustwidth}

\section{Related Works}
\label{sec:related_work}
We organize related prior work along several dimensions: (1) the cognitive process of writing, (2) AI-assisted writing interfaces, and (3) document structure and hierarchical organization tools. By synthesizing insights from these areas, we identify recurring gaps between what cognitive theories of writing suggest and what existing tools provide—namely, the lack of support for hierarchical planning, cognitive offloading, and intelligent text manipulation at scale. These gaps directly motivate the design of TreeWriter.

\subsection{The Cognitive Process of Writing}
Hayes and Flower’s foundational cognitive model conceptualizes writing as a complex cognitive activity characterized by interwoven cycles of planning, text generation, and revision ~\cite{flower}. Empirical studies have shown that in the planning stage, writers form hierarchical goals and often revisit these goals throughout the writing process \cite{flower1980}. In the text generation stage, writers translate ideas from the planning stage into concrete textual content ~\cite{Kellogg}. In the revision stage, writers engage in surface-level editing, such as spelling and paraphrasing, as well as more complex structural changes that revise the text's meaning \cite{stephen, nancy}. During these cognitive processes, semantic, syntactic, and lexical information is stored on the fly within the writer's working memory \cite{Olive2012}, so writers with greater working memory capacity tend to produce more fluent and coherent texts~\cite{Kellogg}.

We draw two conclusions from this cognitive model of writing. First, because writing is a non-linear process, writers often need to revisit and revise intermediate content, such as sketches of arguments, structures, goals, and sub-goals, produced during the earlier stages of the writing process. These cognitive scaffolds guide the writing process, but do not appear in the final text. Therefore, preserving these cognitive scaffolds throughout the writing process is essential, as it enables writers to revisit and reflect on them. Second, because working memory is a bottleneck, interfaces that reduce the load on working memory have a positive impact on how writers plan and revise. Consequently, an effective writing tool must both support the capture and persistence of cognitive scaffolds and provide mechanisms that reduce cognitive load~\cite{khurana2024and}.

\subsection{AI-assisted Writing Systems}
A growing body of work has introduced systems that support human–AI co-writing \cite{reza2025co}. These systems span the stages of the writing process — including planning \cite{siddiqui2025script, visar, luminate}, text generation \cite{luminate, ghostwriter, wordcraft, copoet, masson2025textoshop, visar}, and revision \cite{abscribe, masson2025textoshop, ghostwriter, wordcraft} — and are applied across diverse writing contexts. To facilitate analysis, we classify these works into two categories: (1) those that operate directly within the text to suggest edits or continuations, and (2) those that assist in planning, structure and idea guidance beyond surface-level edits. We refer to the first category as \textit{Inline Editors} and the second as \textit{Conceptual Editors}.

\subsubsection{Inline Editors}
Inline Editors embed LLM-based features directly within the document editing interface, providing users with immediate suggestions and modifications to their selected text content. Earlier works, such as Wordcraft ~\cite{wordcraft} and CoPoet ~\cite{copoet}, enable users to invoke LLMs to generate continuations based on selected text and/or a prompt. Wordcraft also allows for rewriting or elaborating in place within the document. Recent works have provided more sophisticated methods for LLM-assisted interactions within text. TextoShop ~\cite{masson2025textoshop} offers features such as rewriting, extending, shortening, style change, grammar checking, and text merging through a drawing-software-inspired interface. GhostWriter~\cite{ghostwriter} also provides rewriting and extending features, but enhances these features through personalized style and context settings. ABScribe~\cite{abscribe} focuses on the revision stage of writing by providing an interface for generating, storing, and comparing multiple alternative versions of a selected passage adjacent to the original text. This allows the user to explore different phrasings and select the most appropriate option for their context.

Inline approaches face limitations. While these tools provide immediate suggestions for improvement and enable authors to refine or accept AI-generated rewrites on the spot, they typically operate at the sentence or paragraph level without considering broader document structure or hierarchical relationships beyond the user-selected content. Therefore, the common interaction pattern in current inline editors—manually selecting relevant text, inserting prompts, and regenerating content—does not generalize to long-form writing, as it requires repeated modifications that become tedious as the document size grows. Although users can technically select the entire document for LLM modification, they typically prefer fine-grained control over the generated content instead of handing it all to the model~\cite{2022creative, reza2025co}.

\subsubsection{Conceptual Editors.}

In contrast to inline editors, conceptual editors support writing through higher-level structures and interactions in addition to inline text edits, focusing on planning, organization, and creative exploration. To facilitate idea exploration, Luminate~\cite{luminate} utilizes a user-provided writing prompt to generate a collection of drafts, each comprising a particular set of labels belonging to dimensions such as setting or tone. This provides a structured view for exploring alternative drafts, as the user can organize drafts into clusters that share the same labels. Script\&Shift~\cite{siddiqui2025script} offers a layered editing interface to support idea exploration. In Script\&Shift, the user can create drafts that are stored as draggable editor windows called layers within a shared workspace. The system enables quick iterations on drafts by providing layer-level operations such as prompt-based rewriting, inter-layer comparison, child layers, and stacking layers together to visualize the final linear document. VISAR~\cite{visar} supports visual writing planning by providing an interface that displays high-level arguments, discussion points, supporting evidence, and counterarguments as nodes in a tree connected by logical relationships. This allows the user to edit the draft by adding or modifying nodes in the tree. 

While these systems elevate writing beyond direct text manipulation, they often operate within limited levels of abstraction. For instance, VISAR’s abstraction model is tightly coupled to argumentative writing and lacks generality for broader forms of creative or expository writing. Script\&Shift allows adding child layers iteratively; however, its linked layers lack explicit abstraction relations, functioning more as annotations than abstractions, which limits the form of hierarchical structure.
As a result, existing conceptual editors still fall short of supporting representing and operating multi-level abstraction of documents.

\subsection{Document Structure and Organization Tools}
Empirical studies have shown that writers who adopt a structured approach to writing consistently produce texts with better overall quality ~\cite{desmet2011,desmet2012,limpo2018,klein2015}. For instance, students who repeatedly used electronic outlining tools developed clearer text structures and reported lower mental effort during writing \cite{desmet2012}. Similarly, students who plan with a structured outline that embeds the main arguments performed better than those who use an unstructured list of ideas \cite{limpo2018}. When writing tasks are presented as a set of sub-goals, it is also found that fine-grained goals reduced cognitive load and improved reasoning compared to clustered goals \cite{klein2015}.

Consistent with these findings, several existing tools incorporate various features that support a structured approach to writing. Scrivener~\cite{scrivener} provides corkboard and outlining features. \LaTeX-based systems such as Overleaf ~\cite{Overleaf} supports hierarchical section management. However, these tools either incorporate no AI functionality or restrict their assistance to in-line text improvements, lacking mechanisms that leverage inter-section relationships to provide structure-aware AI assistance.

TreeWriter contributes to this landscape by combining the strengths of both inline and conceptual editing paradigms with a tree-based hierarchical interface. In the document trees of TreeWriter, non-leaf nodes preserve cognitive scaffolds such as goals, outlines, and summaries, while the exported content, mostly from leaf nodes, corresponds to the final textual output. This design enables structured planning and reduces cognitive load when switching contexts between planning, text generation, and revision. Unlike existing inline editors that operate on individual text segments, 
TreeWriter features a chat-based, agentic writing assistant that operates on the nodes of the tree. This design enables TreeWriter to automate the ``select–prompt–generate'' interaction pattern across the document by leveraging its agentic capabilities to perform multi-step interactions to load the essential context and generate document content at both high and low levels.

\begin{table}[t]
\centering
\small
\begin{tabular}{lcccccc}
\toprule
 & Wordcraft & Luminate & Textoshop & Script\&Shift & VISAR & \textbf{TreeWriter} \\
 & \cite{wordcraft} & \cite{luminate} & \cite{masson2025textoshop} & \cite{siddiqui2025script} & \cite{visar} & \textbf{(This work)} \\
\midrule
Document structure & Linear & Linear & Linear & Node+Linear & Node+Linear & \textbf{Node+Linear} \\
AI suggestion & Inline & Inline & Inline & Inline & Inline \& Nodes & \textbf{Inline \& Nodes} \\
\midrule
Node relation & -- & -- & -- & Arbitrary & Logic-based & \textbf{Abstract-based} \\
Progressive disclosure & -- & -- & -- & Partial & Partial & \textbf{Yes} \\
\bottomrule
\end{tabular}
\caption{Comparison of AI writing tools. TreeWriter uniquely employs abstract-based node relations, where we prompt the user and agents to make the parent nodes contain the outlines of their child nodes. This enables progressive disclosure for both human and AI agents and therefore supports scalable interaction with long documents.}
\label{tab:ai-compare}
\end{table}

\section{Formative Study}
\label{sec:formative_study}

To better understand the challenges of long-form writing and attitudes toward AI-assisted writing, we conducted a formative study with researchers who regularly produce long documents. Insights from this study directly informed the design of TreeWriter by highlighting pain points in document writing and surfacing user preferences for AI-assisted workflows.

\subsection{Study Design and Participants}

We recruited six participants through academic networks and personal connections (their anonymized information is available in \autoref{app:participants}). Participants included graduate students and postdoctoral researchers who routinely produce documents exceeding ten pages in length, such as research articles, grant proposals, and technical reports. The study was conducted through an online structured survey that combined open-ended interview questions with Likert-scale questions. The complete list of questions is available in \autoref{app:formative_question}.

The interview questions focused on two main areas. The first explored the participants' background and existing practices in academic writing. The participants were asked to describe their typical processes for working on long documents, the most significant challenges they face, and what features they wish for in writing tools to better support this work. The second area covered their perceptions of AI in writing. This section explored participants' prior use of AI tools for writing tasks, the impact those tools had on their workflow, and their perspectives on what constitutes responsible AI support for collaborative writing. Likert-scale questions were aligned with these topics and served as a complementary supplement to the interview questions.

\subsection{Findings}

Analysis of responses revealed three recurring challenges that shaped the design of TreeWriter:

\subsubsection{C1. Structuring Ideas and Transforming Them into Text}
Participants described writing as an iterative cycle of outlining, expanding, and refining. F1 notes that ``I usually start with a one-page draft and ask the AI to expand it into ten pages—it saves about a week of work.'' A typical process (n=4) involved sketching a high-level structure (e.g., a short draft, outline, or mind map), drafting sections in collaboration with AI tools, and then engaging in human-led refinement to improve style, accuracy, and formatting. This iterative loop enabled authors to focus on high-level ideas while offloading more mechanical drafting tasks; however, it also highlighted gaps in tool support for structuring and navigating evolving ideas. Most participants (n=5) agreed or strongly agreed that writing long documents is often frustrating and that they frequently lose track of the overall structure, highlighting the need for writing tools that support high-level organization. For instance, one participant (F4) wished for, ``I would like a more interactive writing tool that can guide me through the document-building process — taking a high-level outline as input and prompting me to fill in details.'' 

\subsubsection{C2. Maintaining Consistency Across Sections}
Ensuring coherence in terminology, argumentation, and narrative flow across sections was a persistent challenge. As one participant (F2) explained, ``It's challenging to correlate the context and keep high-quality writing consistent from beginning to end.'' Participants (n=2) mentioned that maintaining coherence is challenging in long-form documents. Participants also envisioned tools to better preserve coherence across sections, for example, ``a helper that can maintain long context and summarize what's been written'' (F2).

\subsubsection{C3. Verifying Accuracy and Trustworthiness}
Participants expressed concerns over the factual accuracy of using AI-generated text. As one participant warned, ``AI tools sometimes cite fake or incorrect references, which causes a lot of trouble'' (F3).  They reported instances of hallucinated citations, misleading references, or technically inaccurate claims, all of which demanded substantial verification effort. Likert responses indicate participants were generally comfortable using AI to assist writing but showed more caution about trusting AI to draft content autonomously, highlighting the importance of verifiable outputs. As another participant emphasized, ``Responsible AI should at least verify the authenticity and reliability of the references it provides'' (F6).

\subsection{Design Goals}

Guided by insights from our formative study (\autoref{sec:formative_study}), we designed TreeWriter to address the central challenges of long-form writing while ensuring responsible AI support. The resulting design goals map directly to the three challenges identified in the study.

\begin{enumerate}
    \item[\textbf{DG1}:] \label{dg1} \textbf{Support Idea Structuring and Iterative Drafting.}  
    Enable writers to organize their ideas hierarchically and develop them progressively into text. This involves creating outlines, structuring content across multiple levels, and supporting iterative expansion from high-level concepts to detailed paragraphs.

    \item[\textbf{DG2}:] \label{dg2} \textbf{Ensure Consistency Across Sections.}  
    Promote coherence throughout a long document by utilizing the hierarchy. This includes detecting inconsistencies at each level and providing guidance that maintains continuity across sections.
    
    \item[\textbf{DG3}:] \label{dg3} \textbf{Enable Trustworthy and Transparent AI Collaboration.}  
    Ensure that AI support is accurate, reliable, and controllable. This involves presenting AI suggestions transparently, allowing authors to review and revise outputs, and maintaining accountability for changes made with AI assistance.
\end{enumerate}

These goals underpin both the hierarchical editor and its AI features, ensuring that TreeWriter provides helpful and trustworthy support for long-form writing.

\section{TreeWriter}

\begin{figure}
    \centering
    \includegraphics[width=1\linewidth]{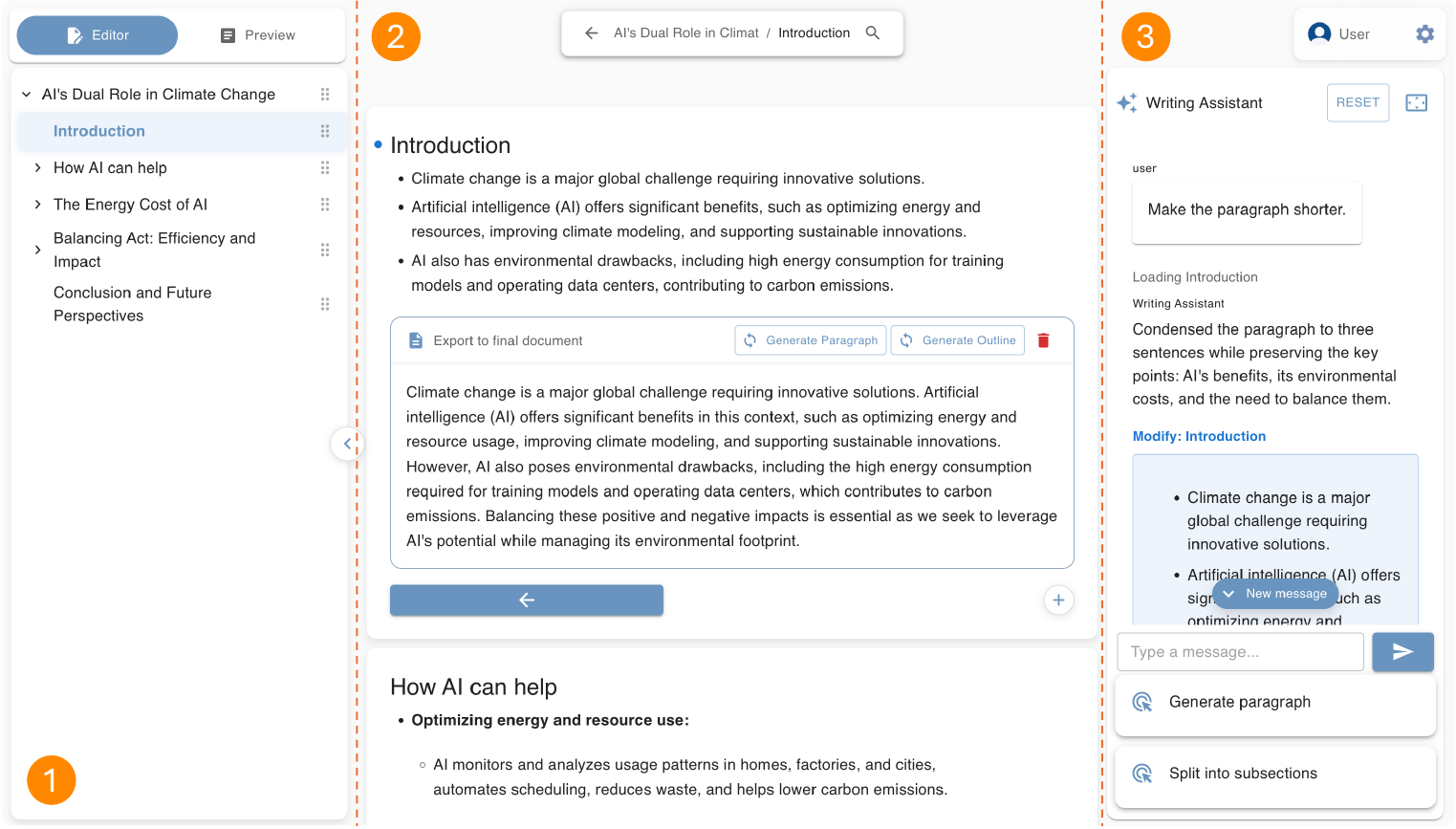}
    \caption{TreeWriter’s interface consists of three columns:
(1) a \textbf{tree navigator} on the left for organizing the document structure, with a view switcher at the top for changing the middle-column view;
(2) the \textbf{middle column}, which supports two complementary modes: a \textbf{tree view} (currently shown) for hierarchical editing by displaying the children of a parent node, and a \textbf{linear view} for previewing the composed text of that node’s subtree. A floating toolbar at the top allows navigation to higher-level nodes and searching within the document;
(3) a chat-based \textbf{writing assistant} and AI editing buttons on the right for AI-assisted document editing.}
    \label{fig:system}
\end{figure}

Based on the design goals, we designed and implemented \textbf{TreeWriter}. TreeWriter's interface is organized into three columns (See \autoref{fig:system}): the left column is a tree navigator where nodes can be reordered through drag-and-drop, the middle column displays document content, and the right column hosts the chat-based writing assistant and AI-powered editing buttons. The middle column supports two complementary views: a \textit{tree view} for structural and conceptual editing, and a \textit{linear view} for previewing the final document. These two views can be switched by the buttons at the top of the left column.

\subsection{Tree View}

\begin{figure}
    \centering
    \includegraphics[width=\linewidth]{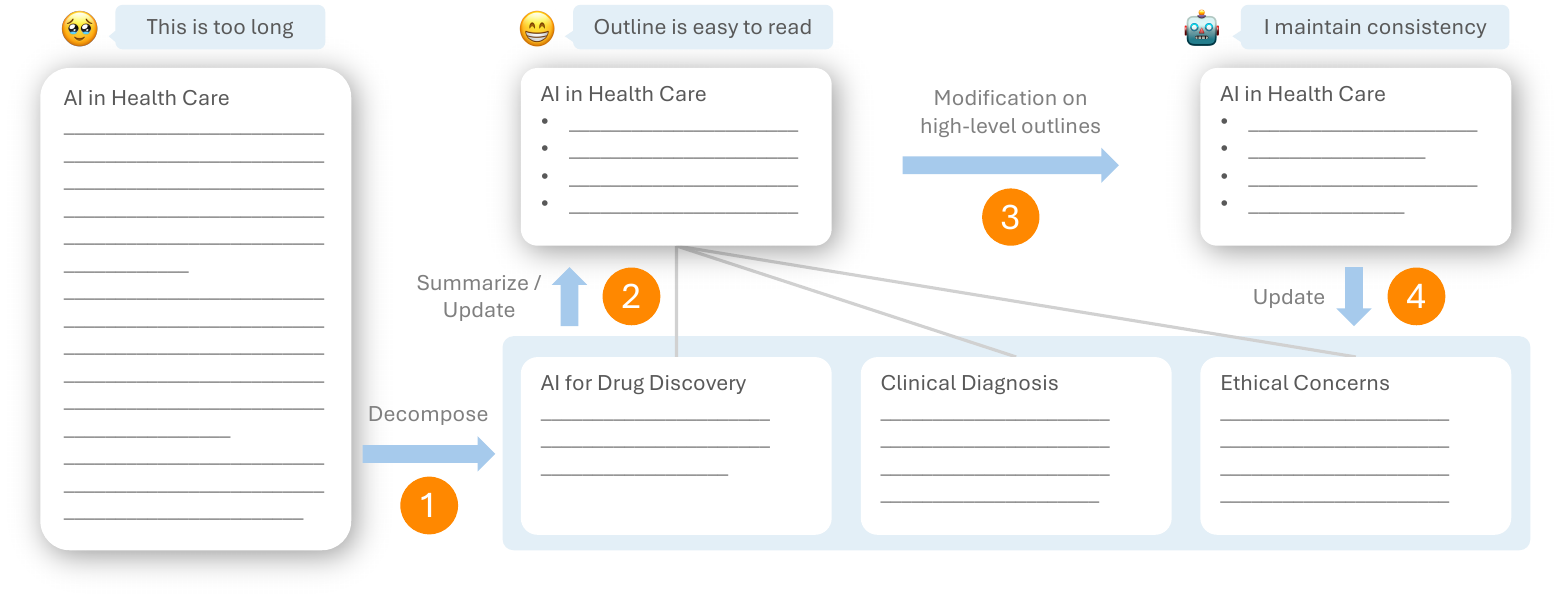}
    \caption{AI-assisted abstraction creation. (1) TreeWriter lets the user freely write within a node, which can grow to any length. Later, the user can split it into reasonable chunks using the \textbf{``Split into subsections''} button, which leverages an LLM to generate child nodes that collectively cover the original text. (2) Once the text has been split, the user can use the \textbf{``Generate outline from children''} button to rewrite the parent node as a concise outline of its children. This ``split\&summarize'' process reduces large nodes and makes the whole tree more reader-friendly. This function can also be used after substantial edits to the children, so the parent content remains in sync with the children. (3) Users can revise a section at a high level by modifying its root node and then use AI to propagate these changes to the subtree. (4) In response to a chat request, TreeWriter's writing assistant can update the child nodes to ensure that the final content reflects the revised outline. This makes it convenient to maintain consistency between the higher-level outline and the lower-level realization during the revision stage of writing. The writing assistant can also be asked to maintain the coherence of the child nodes.}

    \label{fig:decompose}
\end{figure}

In the tree view, the middle column displays the children of a specific parent node. Each node is shown as a card containing a title and editable content. Navigation controls at the bottom of each card enable the user to navigate to the parent, navigate to the node's children, add a new child, or delete the current node. Nodes can also be reordered via drag-and-drop. When the user clicks a node, a blue dot will appear at the top-left corner of the node to show it is ``selected'', which provides context to AI assistance.

This design supports a multi-level hierarchical representation of documents: parent nodes summarize their children, and children elaborate on their parent. Such hierarchical decomposition allows users to iteratively refine high-level ideas into detailed content while maintaining the conceptual structure behind the document.

In order to retain the outlines, TreeWriter does not directly output all the content in the nodes to the final document. Node content connects to the final document in two ways: (1) authors can add an \textit{export block} at the end of a node's content (See \autoref{fig:bidirect-edit}), which will be included in the final document, or (2) if no export block is specified, leaf-node content is exported by default. This flexible mechanism helps users distinguish between structural notes and publishable text, ensuring clarity for collaborators and AI assistance without cluttering the final output.

\subsection{Linear View}

\begin{figure}
    \centering
    \includegraphics[width=1\linewidth]{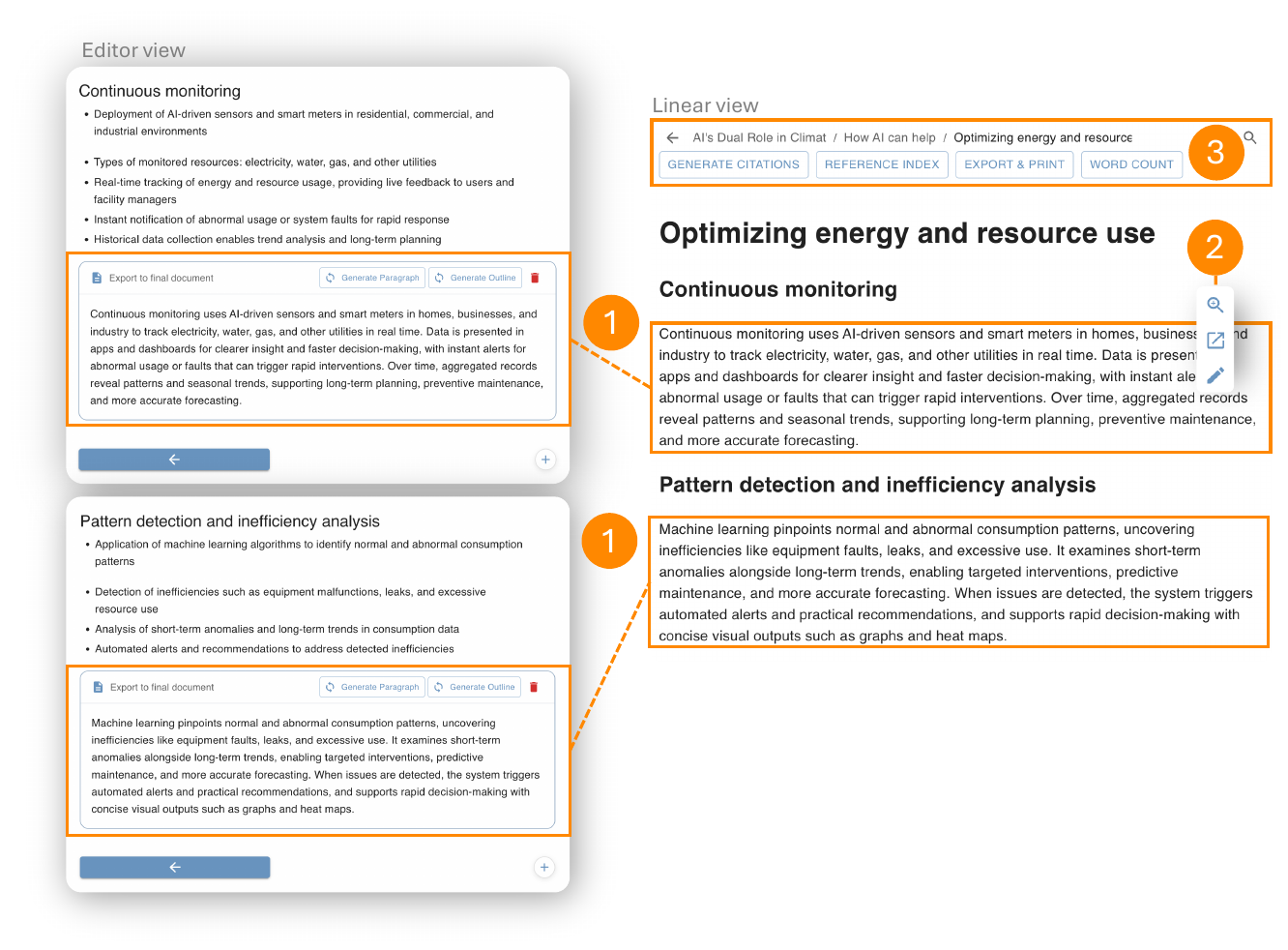}
\caption{Transform from Tree view to Linear view. TreeWriter connects the node-based editor to the final document through the export blocks. Only the content in the export block and the content in a leaf node without an export block appear in the final document. In the linear view, the tree (or a selected subtree) is linearized by a preorder traversal and the content exported from each node is listed in the view. (1) Only the exported content of the nodes is included in the final document. (2) In the linear view, a menu appears when the user clicks on each exported content, which allows the user to focus on the subsection from that node, jump to the corresponding node in the tree view or directly edit that node's content in place. (3) A navigator can be used to expand the scope of the linear view to the ancestors of the current section.}

    \label{fig:linear_view}
\end{figure}

The linear view displays a section of the final document by concatenating exported content from all nodes (See \autoref{fig:linear_view}). Users can scroll through this view to inspect how the final text will appear, and use a toggle button at the top of the right column to switch between tree and linear views.

Within the linear view, each node supports a floating menu which appears when the user clicks. The menu has three buttons: (1) navigate to the corresponding node in the tree view, (2) display only the subtree of the clicked node for focused reading, and (3) toggle edit mode to show and allow the user to directly edit the whole node content, including the parts not exported.

\subsection{Complementary Roles of the Two Views}

In TreeWriter, the tree view serves as the primary interface for writing and organization, while the linear view supports holistic inspection of the evolving manuscript. By combining the tree view and the linear view, TreeWriter reduces cognitive load, preserves structural clarity while maintaining compatibility with traditional formats of documents. 

\subsection{AI-Powered Editing Buttons}

\begin{figure}
    \centering
    \includegraphics[width=\linewidth]{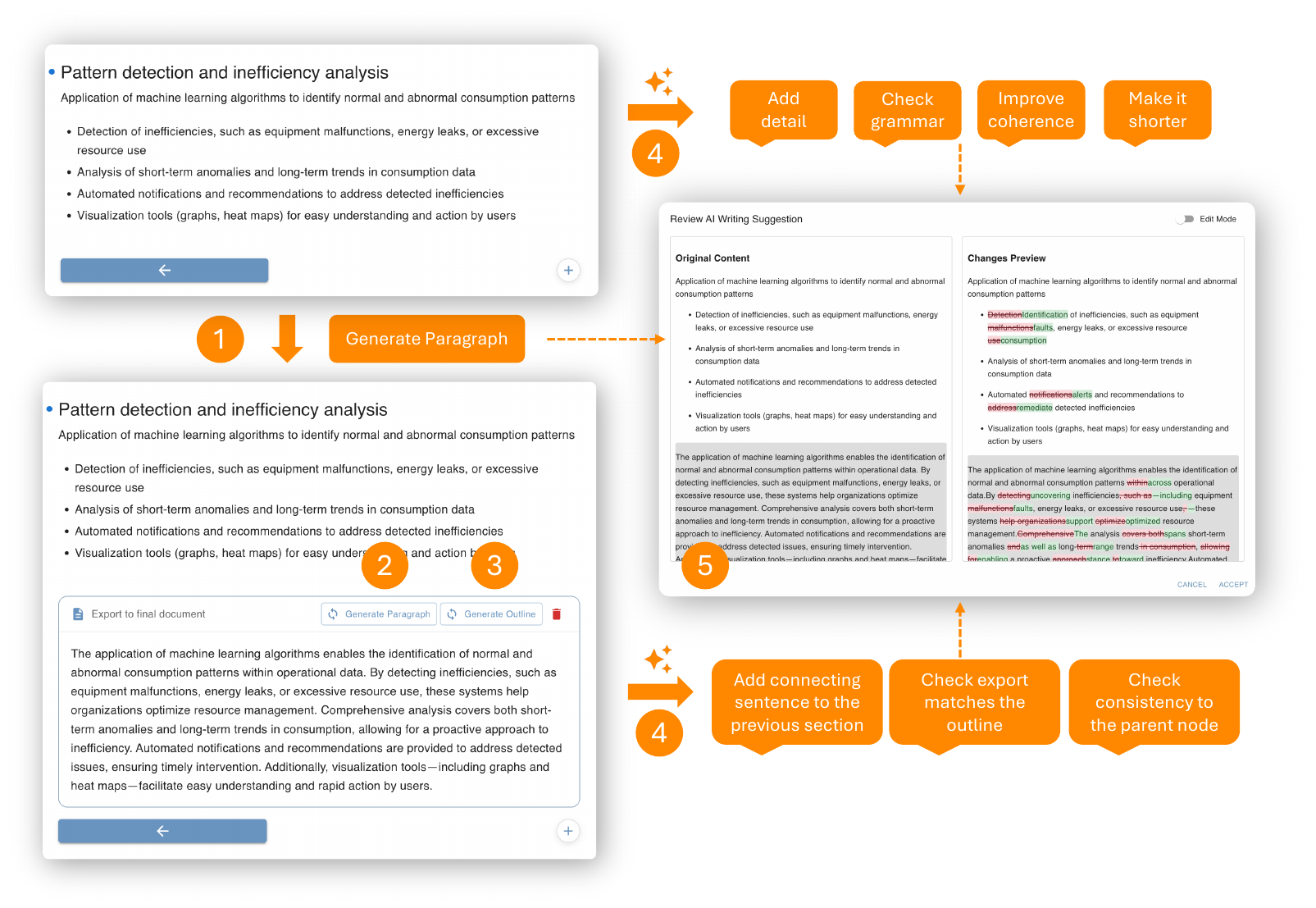}
    \caption{AI-assisted paragraph generation in TreeWriter. Users are encouraged to create and save both the outline and the corresponding prose in the nodes. (1) The user can draft an outline first, then click \textbf{``Generate paragraph''} to create an export block with a draft paragraph ready for inclusion. (2--3) Each export block provides two sync buttons: generating a paragraph from an outline or generating an outline from a paragraph. These features allow flexible editing from either direction while maintaining consistency between the outline and the paragraph. (4) A chat-based writing assistant can refine both outlines and paragraphs by the user's instruction, with awareness of the context of the node in the document. (5) When the writing assistant modifies existing content, a confirmation dialogue with a difference viewer will let the user review, accept, or adjust the proposed changes.}

    \label{fig:bidirect-edit}
\end{figure}

It can be laborious for users to manually maintain the consistency of the hierarchical outlines in the tree. To assist users in building and maintaining the abstraction hierarchy (DG1, DG2), TreeWriter integrates several AI-powered editing buttons, which are displayed at the bottom of the right column.
The editing buttons include:
\begin{itemize}
    \item \textit{Split into subsections}: Breaks down the content of the selected node into several child nodes of it to improve readability and facilitate further elaboration on them.
    \item \textit{Generate outline from children}: Replace the content of the selected node with a summary of the child nodes of it to create abstraction or maintain parent-child consistency.
    \item \textit{Generate paragraph}: Adds an export block to the end of the selected node with a paragraph generated from the outlines in the block.
\end{itemize}

Additionally, within the export block, there are two buttons (See \autoref{fig:bidirect-edit} (2-3)):
\begin{itemize}
    \item \textit{Generate paragraph}: Generates a paragraph based on the outlines in the node.
    \item \textit{Generate outline}: Generates an outline based on the exported paragraph in the node.
\end{itemize}

When an editing button is clicked, the content and context of the node are sent to an LLM to generate the corresponding output. The LLM is prompted to preserve the existing content whenever possible (See \autoref{app:ai_button_prompt} for detailed prompts). After the content is generated, a confirmation dialogue will appear, allowing users to review and edit the generated content before accepting it, reinforcing DG3.

\subsection{Agentic writing assistant on Tree}

\begin{figure}
    \centering
    \includegraphics[width=\linewidth]{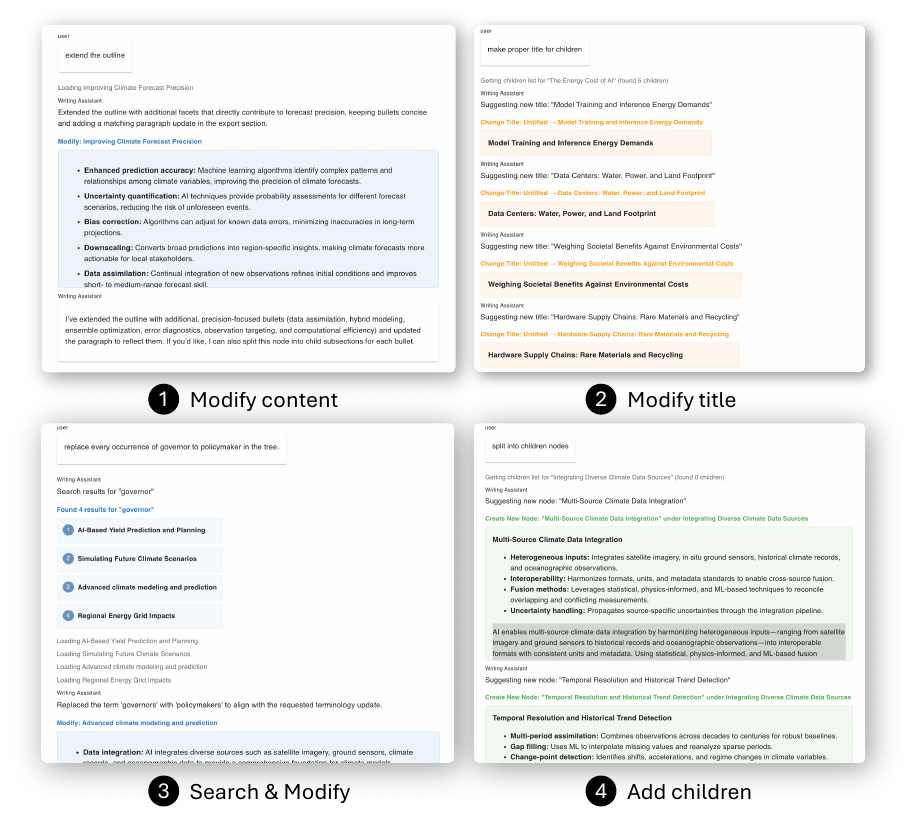}
    \caption{Writing assistant. TreeWriter includes an agentic writing assistant that is aware of the context of the nodes to edit. The assistant can use tools to dynamically load the contents of the node and assist the user in editing the tree. (1,2) The assistant can provide editing suggestions on the content and title of nodes based on the user's requests. Suggestions are displayed as blocks with a blue background in the chat interface; when clicked, a confirmation dialogue appears to help the user confirm the change. (3) The assistant can search for nodes by keywords, enabling it to locate and edit nodes based on the user's request. The search results are displayed in the chat interface, and users can navigate directly to the nodes by clicking. (4) The assistant can also suggest adding new child nodes; the user can accept these suggestions by clicking blocks with a green background.}
    \label{fig:assistant}
\end{figure}

In the right column of the interface, TreeWriter provides a chat interface for users to interact with the LLM-based writing assistant (See \autoref{fig:assistant}). The chat interface allows users to ask questions, seek suggestions, and discuss ideas with the assistant. The chat history is preserved unless the user clicks the reset button, enabling context-aware conversations that build upon previous interactions.

The writing assistant is constructed by assigning proper context and tools to an LLM-based agent.
The list of tools that are available to the agent can be found in \autoref{tab:agent-tools}, which includes making editing suggestions to the user, loading node content and searching nodes by keywords.
The context of the agent includes the title and content of the selected node and its parent. The titles and ID of the sibling and child nodes of the selected node are also presented in the prompt of the agent to enable a dynamic loading of their content based on their ID. 

All this contextual information enables the writing assistant to generate editing suggestions that are not only locally fluent but also globally consistent with the surrounding nodes. The user can request \textbf{localized edits} targeting a single node—for example, asking the assistant to check whether a node’s content aligns with that of its parent, siblings, or children (DG2). The tree structure also supports efficient high-level modifications: after revising a high-level node, the user can instruct the assistant to update its child nodes accordingly, achieving \textbf{high-level conceptual editing} (DG1). Moreover, the user can direct the assistant to focus on a specific concept within the document, prompting it to locate related nodes through keyword search or structural traversal (e.g. beam search) and then make modification suggestions (DG1).


\begin{table}[h!]
\centering
\begin{tabular}{p{0.28\linewidth} p{0.68\linewidth}}
\toprule
\textbf{Tool name} & \textbf{Description} \\
\midrule
\textit{Load Node Content} & Loads a node's content into the context by its ID. \\
\textit{Load Node Children} & Loads the ID and title of the child nodes of a node into the context. \\
\textit{Suggest New Title} & Suggests a new title of a node for the user to review. \\
\textit{Suggest New Content} & Suggests a new version of the content of a node for the user to review. \\
\textit{Suggest New Child} & Suggests a new child to a certain node for the user to review. \\
\textit{Search by Keyword} & Searches for nodes that contain a given keyword from the whole tree. \\
\bottomrule
\end{tabular}
\caption{Available tools in TreeWriter's writing assistant. These tools enable the assistant to modify or add content to the nodes.}
\label{tab:agent-tools}
\end{table}


\subsection{Confirmation and Versioning}

To uphold transparency and trust (DG3), all AI-generated editing suggestions undergo review through an interactive confirmation dialogue before being applied (See \autoref{fig:bidirect-edit}). The dialogue employs a two-column layout that displays the original text on the left and a detailed difference view highlighting proposed AI changes on the right, enabling authors to quickly understand what is going to be modified.

An optional \textit{edit mode} empowers authors to modify AI-generated editing suggestions before acceptance; the difference view updates dynamically once edit mode is deactivated, showing the final changes relative to the original content. All accepted revisions are systematically archived with descriptive labels in a comprehensive version history, enabling easy restoration, side-by-side comparison, and safe experimentation without risk of data loss.


\subsection{Implementation detail}

We used TypeScript, React and Material UI \footnote{https://mui.com/} to develop the user interface. We used the TipTap \footnote{https://tiptap.dev/} framework for building the editor. To enable collaborative editing in TreeWriter, we integrated Yjs \cite{nicolaescu2015yjs}, allowing multiple users to edit simultaneously. On the backend, we employed TypeScript and Express.js. The writing assistant is implemented by the Vercel AI SDK \footnote{https://ai-sdk.dev/} and uses GPT-5 \cite{openai2025introducinggpt5} from OpenAI as the backbone model. The AI-powered editing buttons utilize GPT-4.1 \cite{openai2025introducinggpt4-1} for quicker response to users.

\section{Comparative Lab Study}

We conducted a user study to evaluate the effectiveness of TreeWriter in supporting long-form AI co-writing compared to a baseline system. The study aims to answer the following research questions:

\begin{adjustwidth}{4em}{}
\begin{itemize}
 \item [\textbf{RQ1}:] (Edit existing long document) How does TreeWriter, with its AI-assisted structure and writing assistance, help users edit and reorganize existing long-form documents? In particular, does the AI assistance help users work more efficiently and maintain a clearer sense of structure and consistency across sections compared to baseline tools?

 \item [\textbf{RQ2}:] (Draft from ideas to document) How does TreeWriter support users in transitioning from brainstorming and outlining to drafting coherent prose? In particular, does it help users explore and organize ideas more effectively and perceive greater support from AI assistance compared to baseline tools?

 \item [\textbf{RQ3}:] (Authorial control, transparency, and trust) Does TreeWriter increase authors' perceived and behavioural control over AI-generated content, improve transparency of changes, and reduce verification effort and perceived risk of inaccuracies, resulting in outputs that better reflect the author's intent?
\end{itemize}
\end{adjustwidth}

\subsection{Participants and Procedure}

We recruited 12 participants (P1-P12; 2 women, 10 men) through the mailing lists and Discord channels of student clubs at a research-focused Tier 1 university in North America. Detailed information about the participants can be found in \autoref{app:participants}.
The participants include: five undergraduate students, three master's students, and four PhD students, with representation from a diverse set of disciplines (computer science, law, and chemistry). 
All participants reported prior experience with co-writing using AI. Each study session lasted approximately 2.5 hours, and the participants received compensation of \$50 CAD.

\subsection{Study Design}
Each study session consisted of two main tasks: \textbf{Article Modification} and \textbf{Creative Writing}. In the article modification task, the participant is provided with a 4,000-word article on AI's implications for the world and is asked to complete six editing sub-tasks on it. 
In the creative writing task, the participant is given a topic on AI and is asked to write an 800-word essay. We adopted a within-subjects design, in which each participant completed the tasks using both TreeWriter and a control (Google Docs + Gemini). 

The tasks were chosen to fit within the time constraints of each user study session and to hold real-world value. They were modelled after common university-level writing assignments, and the topics were selected to ensure that an average participant would have a moderate amount of familiarity with the subject matter.

The user study session was conducted as follows (detailed procedure can be found in \autoref{app:study1}):
\begin{enumerate*}
\item Introduction to the study, consent, and setup recording.
\item A random tool was set up and introduced through an introductory video.
\item The participants were then asked to complete the article modification task on an article, followed by the creative writing task on a topic. After each task, the participant filled out a post-task questionnaire. This sequence was designed to minimize the cognitive overhead associated with switching between editing and generative modes. 
\item The participant then completed the same tasks on the other tool, article, and topic.
\end{enumerate*}

\subsection{Control} We chose Google Docs with Gemini \cite{google_docs}, a widely adopted linear document editor that integrates AI functionalities as the baseline system (control group). Gemini offers a chatbot-style interface comparable to TreeWriter, leveraging the document as context to provide writing suggestions. It allows participants to insert AI-generated text directly into the document and to revise selected text through natural language instructions.
More details on this can be found here \cite{google_docs_gemini_collab, google_docs_gemini_write}.

\subsection{Measures}
In the post-task questionnaire, participants rated their agreement with seven statements following the article modification task and five statements following the creative writing task. All questions were rated on a 7-point Likert scale (1 = Strongly Disagree, 7 = Strongly Agree). For the article modification task, participants evaluated the helpfulness of AI features, the system’s support for breaking sections into subsections, control over edits, creating summaries from detailed content, expanding outline points into paragraphs, propagating conceptual changes, and obtaining a clear structural overview. For the creative writing task, participants assessed the helpfulness of AI features, support for building and refining document structure, assistance in exploring and developing initial ideas, effectiveness in expanding outline points into paragraphs, and overall productivity. Detailed questions can be found in \autoref{app:study1-questions}.

To complement our custom questionnaire, we administered the Creativity Support Index (CSI)~\cite{cherry2014quantifying} in both conditions. The CSI evaluates six dimensions: Exploration, Expressiveness, Immersion, Enjoyment, Results Worth Effort, and Collaboration. Because our tasks were single-author and non-collaborative, we omitted the Collaboration dimension. 
Participants rated each dimension on a 10-point Likert scale, and we computed per-dimension scores and an overall CSI score following the published procedure. CSI provided a validated, tool-agnostic measure of perceived creativity support, spanning ideation and expression (Exploration, Expressiveness), engagement (Immersion, Enjoyment), and efficiency (Results Worth Effort).

After completing two tasks, we also administered a NASA-TLX questionnaire~\cite{hart1988development} to assess cognitive load in both tasks under both conditions. This provides a standardized measure of mental demand, physical demand, temporal demand, performance, effort, and frustration associated with the writing tasks.

\subsection{Results}

\subsubsection{How TreeWriter helps editing long documents (RQ1)}

\begin{figure}
    \centering
    \includegraphics[width=\linewidth]{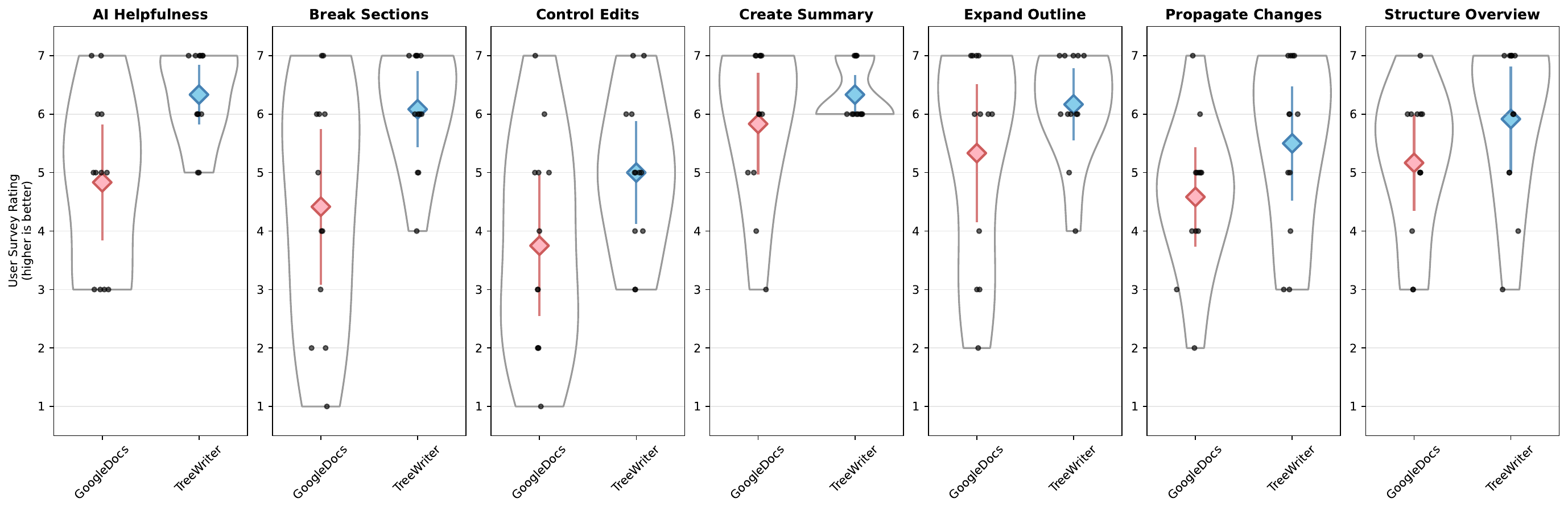}
    \caption{Participant ratings of the \textit{modification experience} in the modification task, comparing TreeWriter and Google Docs across seven dimensions: helpfulness of AI features, breaking sections into subsections, control over edits, creating summaries from detailed content, expanding outline points into paragraphs,  propagating conceptual changes, obtaining a clear structural overview. Ratings were given on a 7-point Likert scale (higher is better). Results show that TreeWriter consistently achieved higher scores with smaller variation, indicating a more effective and consistent AI-assisted editing experience.}
    \label{fig:modification_experience}
\end{figure}

Quantitative results from Task 1 indicate that participants generally found both editors helpful for AI-assisted editing, with TreeWriter achieving consistently higher scores on most questionnaire items (See \autoref{fig:modification_experience}). TreeWriter outperformed Google Docs in ratings for \textit{AI helpfulness} (6.33 vs. 4.83), \textit{breaking sections} (6.08 vs. 4.42), \textit{controlling edits} (5.0 vs. 3.75), \textit{creating summaries} (6.33 vs. 5.83), \textit{expanding outlines} (6.17 vs. 5.33), \textit{propagating changes} (5.5 vs. 4.58), and \textit{providing a structural overview} (5.92 vs. 5.17). These results suggest that TreeWriter offers a more effective AI-assisted editing experience, with higher ratings and less or similar variation across participants.

Qualitative feedback highlighted two main strengths. First, participants emphasized that TreeWriter's hierarchical structure improved organization and navigation, particularly for managing large and complex documents. As one participant explained, ``The hierarchical structure presents a more intuitive view compared to Google Docs, as a tree structure is more organized than just a linear structure'' (P11). The response from another participant (P1) reinforced this advantage: ``[TreeWriter is] easy to manage large intricate documents and easy to navigate - no need to keep scrolling.'' 

Second, participants appreciated that TreeWriter's writing assistant could operate across multiple levels of granularity, supporting both local refinements and global restructuring. One participant noted, ``The structural nodes allowed me to control the scope and granularity of changes wanted.'' Another highlighted the global utility: ``The overall AI Assistant is smart enough to look over the tree for all prompts, allowing both granular and coarse editing together.'' Together, these features enabled participants to not only polish sentences but also reshape sections and maintain coherence across the document.

During the study, we also observed a strong tendency for participants to rely on the writing assistant to complete tasks without manually navigating the document. The tendency is observed in both TreeWriter and the control system. A common strategy was to use a search to locate a section, then delegate most of the editing to the writing assistant. In tasks without a clear edit location, participants often prompted the assistant to search the tree itself; for instance, one participant (P9) directly instructed the writing assistant to scan from the root node, which successfully identified the correct location without requiring manual navigation. This demonstrates how our work effectively leverages the tree structure and node-level outlines to target edits accurately.
\begin{figure}
    \centering
    \includegraphics[width=\linewidth]{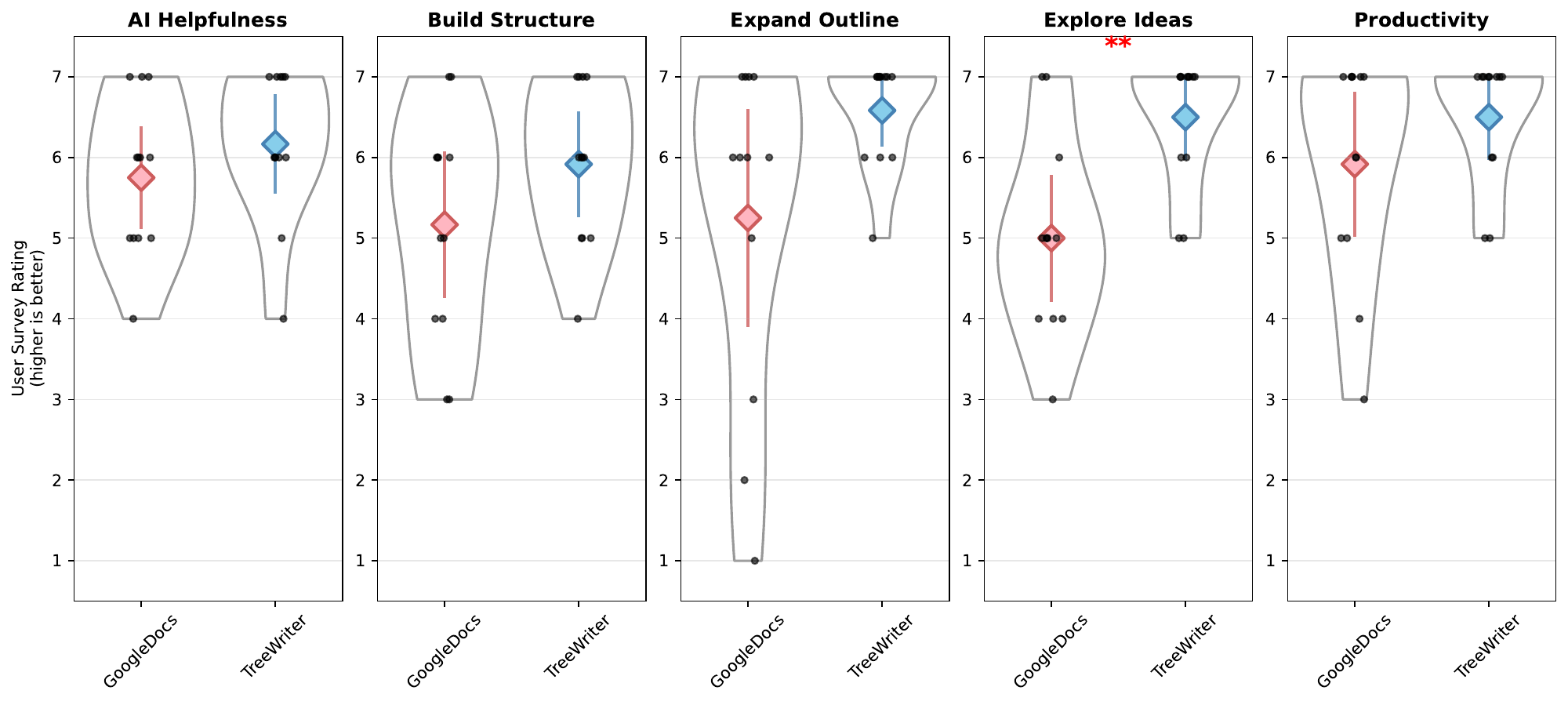}
\caption{Participant ratings of the \textit{tool usefulness} in the creative writing task, comparing TreeWriter and Google Docs across five dimensions: AI helpfulness, building and refining document structure, expanding outline points into paragraphs, exploring and developing initial ideas, and overall productivity. Ratings were on a 7-point Likert scale (higher is better). TreeWriter consistently received higher ratings, indicating stronger support for participants. Notably, the rating for \textit{exploring and developing initial ideas} (highlighted by red stars) remained statistically significant after applying the within-category Benjamini–Hochberg correction for multiple comparisons (FDR < 0.05, q = 0.01; See \autoref{app:study1-result} for details).}
    \label{fig:creative_tool_usefulness_violin_comparison}
\end{figure}
However, TreeWriter also presented drawbacks. While many praised the tree view, some participants found it overly complex for short documents and unfamiliar compared to Google Docs' linear interface. As one participant put it, ``The hierarchical organization is too complex when the total text document is small, and I need to keep clicking to expand'' (P1). Another remarked, ``TreeWriter's interface requires extra time to learn and may feel less intuitive at first'' (P5). In addition, a few participants (n=2) noted occasional AI hallucinations, which undermined trust in the automation.

\subsubsection{How TreeWriter helps draft new documents (RQ2)}

TreeWriter was frequently described as effective for initiating new writing projects, helping participants break down ideas, create outlines, and structure documents before drafting. Participants emphasized its utility for early-stage writing: ``It was quicker and easier to draft up a basic outline,'' and ``This makes organizing the section and subsections easy before writing any text.'' Others highlighted its value for larger texts, noting that it helped them ``organize the sections as I developed my writing ideas further.''

At the same time, participants reported challenges with AI-generated text, citing hallucinations, redundancy, and wrong word counts. As one participant remarked, ``The AI may hallucinate the ideas presented… for example, I was talking about plagiarism, but TreeWriter's version reframed it as cheating'' (P12). Such issues might limit trust in the system despite its structural benefits. However, it is worth pointing out that the participant was able to quickly identify such an inconsistency when using TreeWriter, which might, in fact, prevent such an inconsistency from being carried to the later stage.

\begin{figure}
    \centering
    \includegraphics[width=\linewidth]{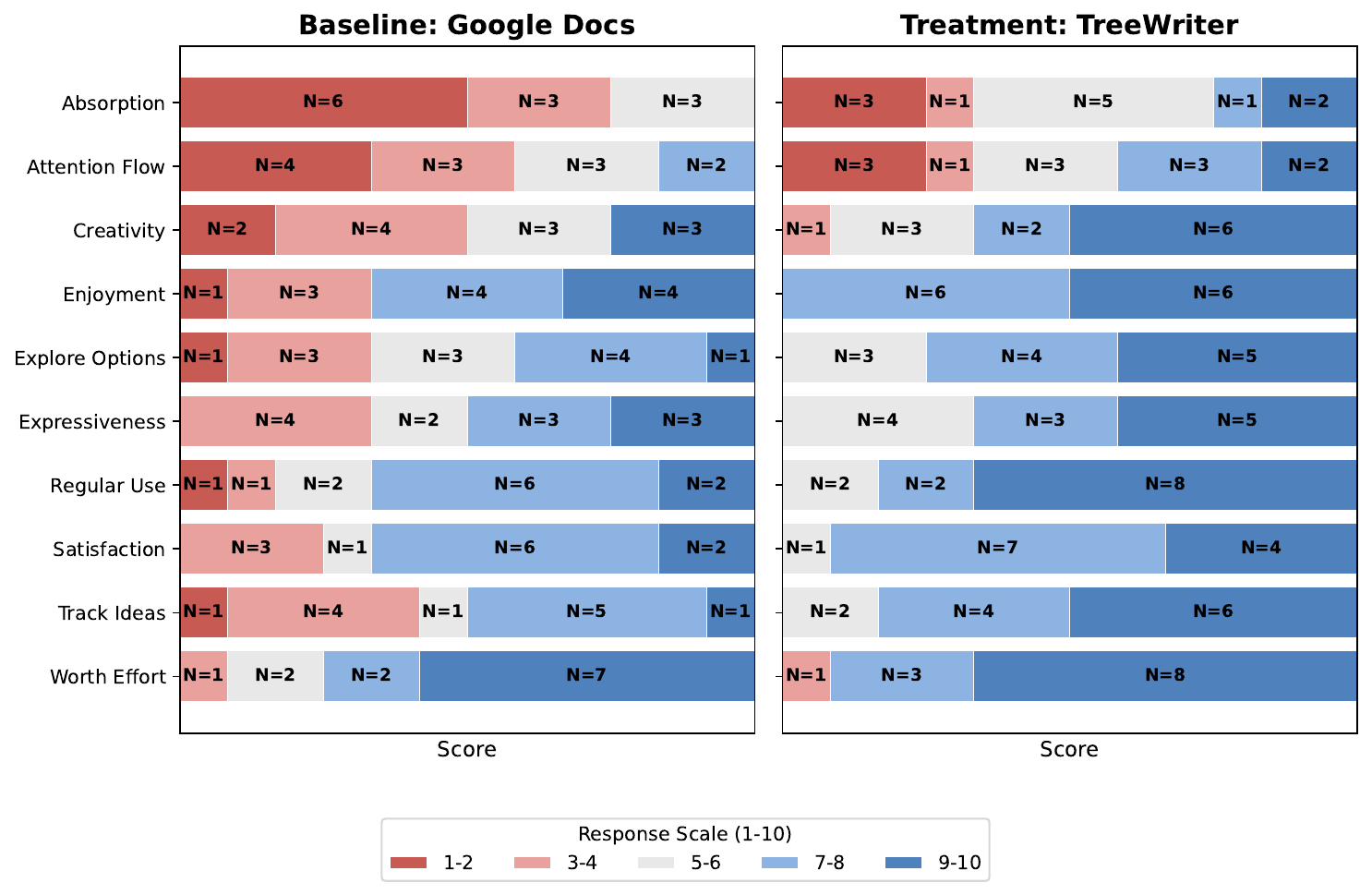}
\caption{Creativity Support Index (CSI) \cite{cherry2014quantifying} ratings (1–10 Likert-scale, higher values indicate better support) comparing TreeWriter and Google Docs. TreeWriter outperformed Google Docs on most dimensions, with gains in creativity, enjoyment, and tracking ideas.}

    \label{fig:CSI_likert_comparison}
\end{figure}

Quantitative results aligned with these perceptions. In the Creativity Support Index (CSI), TreeWriter outperformed Google Docs across nearly all dimensions, with gains in creativity, enjoyment, and idea tracking (See \autoref{fig:CSI_likert_comparison}). Usefulness ratings showed similar patterns: TreeWriter was considered to have more helpful AI and was more helpful for building structure, expanding paragraphs, and presented a significant advantage in exploring and developing ideas (See \autoref{fig:creative_tool_usefulness_violin_comparison}).

Overall, TreeWriter provided more consistent and reliable support for drafting new documents, particularly in outlining and idea development, though concerns remain about the reliability of AI-generated content.

\subsubsection{How TreeWriter helps the user gain control of the generated document (RQ3)}

TreeWriter's fine-grained, node-based editing offered observability that participants valued, especially in comparison to the baseline editor. Several participants reported frustration with the baseline's insertion mechanics, where AI-generated content was sometimes inserted at incorrect locations or with formatting errors, such as entire paragraphs being promoted to headings. A participant (P9) directly asks, ``Can I use an external diff tool?'' when using the baseline.
In contrast, TreeWriter's explicit scope for AI intervention, confined to tree nodes, helped participants review and manage AI edits more deliberately. One participant (P10) reflected, ``The highlighted changes helped me visually compare against different versions, helping me reduce the time and efforts needed to make the needed edits,'' summarizing a general feeling of increased control (See \autoref{fig:modification_experience}).

Nonetheless, TreeWriter did not alleviate a common tendency among participants to confirm AI suggestions with minimal scrutiny, which is also reported in \cite{khurana2024and}. Despite the presence of confirmation dialogues before applying changes, many participants noted that they often accepted suggestions without close inspection, highlighting an ongoing challenge in AI-assisted writing interfaces regarding oversight and critical review.

\section{Study 2: Field Deployment Study}

To evaluate TreeWriter in real-world scenarios, we recruited eight participants (L1–L8; see \autoref{app:participants} for details). Two participants were PhD students in Computer Science, one was a PhD student in Chemistry, one was a Postdoctoral researcher in Computer Science, and the remaining four were Postdoctoral researchers in Chemistry. While we only report their primary fields to preserve anonymity, many conduct research across a different combination of subfields including quantum computing, autonomous discovery, and cheminformatics, making a diverse range of academic perspectives. The eight participants used the system for ideating and drafting a 25,000-word review paper and two research grants (1,900 and 2,500 words) over a two-month period in a collaborative environment.
To match the functionality of common collaborative tools such as GitHub and Google Docs, we implemented inline comments and an integrated issue-tracking system, allowing participants to assign writing tasks to specific nodes and collaborators within the document trees.
Two authors of this work are also involved in the collaborative writing, but they are not counted as participants.
In the study, the participants are not accessible to the writing assistant, and the AI features they use are the AI editing buttons.
After the study, the participants are asked to fill out a feedback form with Likert-scale questions and open-ended interview questions. The detailed questions and results can be found in \autoref{app:study2}.

\subsection{Results}

\subsubsection{Finding 1: Hierarchical Organization Supports Collaboration}
We found that the hierarchical structure helped manage the complexity of collaborative writing. 
Several participants (n=4) highlighted the ease of delegation: ``The subsections can be easily assigned to different collaborators, although this can also be achieved with sectioning in Word.'' 
The result of Likert-scale questions corroborates this point. The question ``The tree structure makes it easy to divide writing tasks and assign responsibility for different sections to team members'' received the highest score among the ratings (M=6.5).
Some participants (n=2) noted that the tree structure helps them focus on their assigned section. L5 notes that ``When working collaboratively, many people can work independently on their assigned sections without interfering with the work of other people.''
L1 remarked that the approach mirrors existing organizational logic: ``I think company roles would fit here nicely, as they tend to have a tree-like structure.'' 

\subsubsection{Finding 2: AI Integration Facilitates Drafting but Raises Quality Concerns}
TreeWriter's integration with LLMs was seen as valuable for brainstorming and early drafting. 
Participants appreciated that they could ``generate text from a set of bullet points'' or receive ``AI suggestions directly in the text,'' describing this as complementary to their existing drafting styles. 
For instance, L2 noted: ``It's nice because it complements my original drafting style — draft in bullet points first before converting to paragraphs.'' 
Yet concerns emerged around the quality and usability of AI-generated \prose. 
For example, L2 observed that the paragraphs ``read fine at first glance but upon closer scrutiny [were] overly verbose with lots of repetitive ideas.'' At the same time, L1 described ``writing redundancies when writing parent node summary, and then child node text.'' 
The lack of clear differentiation between human and AI-generated content further complicated collaboration: ``It's kind of hard to distinguish between human-written content and AI-generated content if the section is not written by me'' (L2).

\subsubsection{Finding 3: Limited Integration Remains a Barrier}
The main limitation of TreeWriter reported by deployment study participants was its integration with existing workflows. Although TreeWriter provided value during the drafting stage, its limited integration with established tools hindered adoption for later stages of writing. 
Participants noted that ``the content has to be exported to a traditional tool in the end, like Word or \LaTeX, to comply with journal guidelines,'' which introduced extra steps into the workflow. 
Technical writing was particularly affected: ``In my work, I use a lot of equations and reference equations using \LaTeX, [but] TreeWriter had the disadvantage that it was harder to do these label-reference usages.'' 
Others pointed to missing features such as figure placement and reference management: ``No compatibility with reference management software like Zotero,'' and ``No method to have figure captions that follow their figure around.'' 
Despite these barriers, participants saw potential if interoperability were improved. As one remarked, ``It's a really interesting idea that can help us design hierarchical plans in a collaborative manner, using AI — which is highly applicable in many fields of knowledge.''

\section{Discussion and Design Recommendations}

\subsection{Separating Content and Ideas to Enable New Interaction}
 
TreeWriter provides a range of potential benefits by separating a document’s final content from its underlying idea structure. For example, non-native English speakers can draft ideas in their preferred language and leverage TreeWriter to produce coherent English \prose, lowering linguistic barriers in document writing. Its hierarchical navigation also enhances accessibility: screen reader users can explore documents at different levels of granularity, reducing cognitive effort when engaging with long or complex texts, while making the content more accessible \cite{lee2020itoc}. Beyond individual use cases, the tree structure facilitates cross-domain knowledge sharing through reusable structural patterns or adaptable templates that capture high-level ideas \cite{gentner1983structure}. Moreover, this representation supports more structured collaboration and feedback, allowing reviewers to write comments directly on specific nodes, which provides hierarchy, enables precise, context-aware suggestions, and simplifies revision tracking.

\subsection{Enhancing Human-AI Collaboration in Long-Form Writing}

Our study suggests that tree-structured representations can help users draft and edit documents more efficiently, and participants appreciate TreeWriter as a more helpful writing assistant. However, LLMs still face limitations when handling long-form documents. Although these models are designed to process extensive contexts, their performance often deteriorates as input length increases \cite{liu2023lost, li2024long, li2023long}. Similar challenges arise in agentic systems, where long tool lists or extended action histories can significantly reduce reliability \cite{yao2024tau, wang2024appbench, barres2025tau}. To address these issues, graph-based \cite{Edge2024FromLT, li-etal-2024-graphreader} and tree-based \cite{sarthi2024raptor, chen2023walkingmemorymazecontext, rezazadeh2025from, cao2025automating} memory structures have been proposed for LLM-based agents. 
For example, GraphReader \cite{li-etal-2024-graphreader} transforms long text into a graph for an agent to explore. 
In Raptor \cite{sarthi2024raptor}, the authors proved that a hierarchical structure can also help LLMs understand long text better.

TreeWriter applies these insights to human-AI writing collaboration. Its node-based tree structure lets users organize ideas hierarchically, a format that LLMs can reason with more effectively. This design supports flexible collaboration: users can delegate detailed writing tasks to the AI and focus on high-level planning, or the AI can guide users in developing the tree structure as a “writing director.” By aligning the document’s conceptual structure with the writing process, TreeWriter not only helps users manage long-form content but also fosters a more interactive and productive human-AI partnership.

\subsection{Ethical Use of AI in Writing: From Human Originality to AI Interpretability}

As AI systems become increasingly capable of generating complex, coherent texts, the ethical foundations of authorship and originality in writing are being re-evaluated \cite{fritz2025understanding}. Traditionally, ethical writing has emphasized human originality. However, when AI begins to contribute to reasoning and composition meaningfully, the ethical focus must expand \cite{mittelstadt2021interpretability}: human originality remains valuable, it also becomes important that the contributions by AI are interpretable and accountable \cite{ai2023artificial,siddiqui2025draftmarks,reza2025co,ramachandram2025}. As AI models gain generative autonomy, authors can no longer reliably trace how specific ideas or reasoning paths emerge \cite{turpin2023language}. Without visibility into the underlying conceptual structure, authors risk losing epistemic control, raising concerns about accountability, trustworthiness, and alignment with human intentions—factors that are especially critical in domain-specific contexts \cite{ennab2022designing,anakok2025interpretability,fan2021interpretability}. The ethics of AI-assisted writing thus evolves toward supporting transparency: the ability to understand, evaluate, and guide the reasoning embedded in AI-generated text\cite{khalifa2024}.

TreeWriter contributes to the solution of this emerging ethical challenge by externalizing the conceptual structure of writing. Its hierarchical tree representation allows both humans and AI to operate within an explicit framework for abstraction, revealing the high-level ideas that underlie the text. By making the underlying document structure explicit, TreeWriter allows authors to see how AI-generated content connects to higher-level ideas and intentions. This visibility helps users identify where AI reasoning may deviate from their goals, give focused feedback on specific sections, and ensure that AI contributions remain consistent with the document’s conceptual design. In doing so, it not only preserves authorial control but also paves the way for ethically integrating fully automated AI researchers and writers into the academic ecosystem, where interpretability and accountability must remain central values \cite{bano2023investigating, Resnik2024, khalifa2024}.

\subsection{Limitations of the studies}

Both the formative and comparative user studies in this work were limited in scope. The formative study involved only six participants, while the comparative lab study included 12 students from a single university. Although participants represented a range of disciplines, the small sample sizes and the predominance of student participants limit the generalizability of the findings to broader professional contexts. The limited number of participants also prevents us from obtaining more statistically significant results. Furthermore, the study tasks were constrained to fit within a two-hour session, which may have encouraged participants to prioritize task completion over naturalistic writing behaviours. The subsequent field deployment study helped mitigate this issue by observing use in more realistic settings. However, this deployment was still guided to some extent by the authors, which may have potentially influenced adoption and usage patterns. Future work should involve more independent deployments to better capture authentic user behaviours and long-term engagement without expert intervention.

\section{Conclusion}

This work introduces \textbf{TreeWriter}, an AI-assisted writing system designed to support long-form writing through a hierarchical structure. By integrating LLMs with a tree-based document structure, TreeWriter enables authors to create, navigate through and edit multiple levels of abstraction—connecting ideas, outlines, and prose within a unified interface. Our evaluation, combining controlled in-lab experiments and field deployments, demonstrates that hierarchical document organization combined with AI assistance can effectively support the challenges of long-form writing. Beyond TreeWriter itself, this work highlights the importance of designing AI writing tools that externalize and operate upon a document’s conceptual structure, enabling writers to manage complexity through progressive abstraction rather than linear expansion. We further conclude that the future of AI-assisted writing lies not only in automating tasks but in amplifying users’ understanding of complex documents.

\section*{Acknowledgment}
The authors would like to acknowledge valuable discussions with Varinia Bernales.
A.A.-G. thanks Anders G. Fr{\o}seth, for his generous support. A.A.-G. also acknowledge the generous support of Natural Resources Canada and the Canada 150 Research Chairs program. This research is part of the University of Toronto’s Acceleration Consortium, which receives funding from the Canada First Research Excellence Fund (CFREF) via CFREF-2022-00042.

\bibliographystyle{ACM-Reference-Format}
\bibliography{main} 

@inproceedings{siddiqui2025script,
  title={Script\&Shift: A layered interface paradigm for integrating content development and rhetorical strategy with llm writing assistants},
  author={Siddiqui, Momin N and Pea, Roy D and Subramonyam, Hari},
  booktitle={Proceedings of the 2025 CHI Conference on Human Factors in Computing Systems},
  pages={1--19},
  year={2025}
}

@inproceedings{masson2025textoshop,
  title={Textoshop: Interactions Inspired by Drawing Software to Facilitate Text Editing},
  author={Masson, Damien and Kim, Young-Ho and Chevalier, Fanny},
  booktitle={Proceedings of the 2025 CHI Conference on Human Factors in Computing Systems},
  pages={1--14},
  year={2025}
}

@article{zhang2025treereader,
  title={TreeReader: A Hierarchical Academic Paper Reader Powered by Language Models},
  author={Zhang, Zijian and Chen, Pan and Du, Fangshi and Ye, Runlong and Huang, Oliver and Liut, Michael and Aspuru-Guzik, Al{\'a}n},
  journal={arXiv preprint arXiv:2507.18945},
  year={2025}
}

@inproceedings{wordcraft,
author = {Yuan, Ann and Coenen, Andy and Reif, Emily and Ippolito, Daphne},
title = {Wordcraft: Story Writing With Large Language Models},
year = {2022},
isbn = {9781450391443},
publisher = {Association for Computing Machinery},
address = {New York, NY, USA},
url = {https://doi.org/10.1145/3490099.3511105},
doi = {10.1145/3490099.3511105},
booktitle = {Proceedings of the 27th International Conference on Intelligent User Interfaces},
pages = {841–852},
numpages = {12},
location = {Helsinki, Finland},
series = {IUI '22}
}

@inproceedings{copoet,
    title = "\textit{Help me write a poem}: Instruction Tuning as a Vehicle for Collaborative Poetry Writing",
    author = "Chakrabarty, Tuhin  and
      Padmakumar, Vishakh  and
      He, He",
    editor = "Goldberg, Yoav  and
      Kozareva, Zornitsa  and
      Zhang, Yue",
    booktitle = "Proceedings of the 2022 Conference on Empirical Methods in Natural Language Processing",
    month = dec,
    year = "2022",
    address = "Abu Dhabi, United Arab Emirates",
    publisher = "Association for Computational Linguistics",
    url = "https://aclanthology.org/2022.emnlp-main.460/",
    doi = "10.18653/v1/2022.emnlp-main.460",
    pages = "6848--6863"
}

@inproceedings{abscribe,
author = {Reza, Mohi and Laundry, Nathan M and Musabirov, Ilya and Dushniku, Peter and Yu, Zhi Yuan “Michael” and Mittal, Kashish and Grossman, Tovi and Liut, Michael and Kuzminykh, Anastasia and Williams, Joseph Jay},
title = {ABScribe: Rapid Exploration \& Organization of Multiple Writing Variations in Human-AI Co-Writing Tasks using Large Language Models},
year = {2024},
isbn = {9798400703300},
publisher = {Association for Computing Machinery},
address = {New York, NY, USA},
url = {https://doi.org/10.1145/3613904.3641899},
doi = {10.1145/3613904.3641899},
booktitle = {Proceedings of the 2024 CHI Conference on Human Factors in Computing Systems},
articleno = {1042},
numpages = {18},
location = {Honolulu, HI, USA},
series = {CHI '24}
}

@inproceedings{luminate,
author = {Suh, Sangho and Chen, Meng and Min, Bryan and Li, Toby Jia-Jun and Xia, Haijun},
title = {Luminate: Structured Generation and Exploration of Design Space with Large Language Models for Human-AI Co-Creation},
year = {2024},
isbn = {9798400703300},
publisher = {Association for Computing Machinery},
address = {New York, NY, USA},
url = {https://doi.org/10.1145/3613904.3642400},
doi = {10.1145/3613904.3642400},
booktitle = {Proceedings of the 2024 CHI Conference on Human Factors in Computing Systems},
articleno = {644},
numpages = {26},
location = {Honolulu, HI, USA},
series = {CHI '24}
}

@article{ghostwriter,
  title={GhostWriter: Augmenting Collaborative Human-AI Writing Experiences Through Personalization and Agency},
  author={Catherine Yeh and Gonzalo A. Ramos and Rachel Ng and Andy Huntington and Richard Banks},
  journal={ArXiv},
  year={2024},
  volume={abs/2402.08855},
  url={https://api.semanticscholar.org/CorpusID:267658049}
}

@inproceedings{visar,
author = {Zhang, Zheng and Gao, Jie and Dhaliwal, Ranjodh Singh and Li, Toby Jia-Jun},
title = {VISAR: A Human-AI Argumentative Writing Assistant with Visual Programming and Rapid Draft Prototyping},
year = {2023},
isbn = {9798400701320},
publisher = {Association for Computing Machinery},
address = {New York, NY, USA},
url = {https://doi.org/10.1145/3586183.3606800},
doi = {10.1145/3586183.3606800},
booktitle = {Proceedings of the 36th Annual ACM Symposium on User Interface Software and Technology},
articleno = {5},
numpages = {30},
location = {San Francisco, CA, USA},
series = {UIST '23}
}

@inproceedings{nicolaescu2015yjs,
  title={Yjs: A framework for near real-time p2p shared editing on arbitrary data types},
  author={Nicolaescu, Petru and Jahns, Kevin and Derntl, Michael and Klamma, Ralf},
  booktitle={International Conference on Web Engineering},
  pages={675--678},
  year={2015},
  organization={Springer}
}

@incollection{hart1988development,
  title={Development of NASA-TLX (Task Load Index): Results of empirical and theoretical research},
  author={Hart, Sandra G and Staveland, Lowell E},
  booktitle={Advances in psychology},
  volume={52},
  pages={139--183},
  year={1988},
  publisher={Elsevier}
}

@article{cherry2014quantifying,
  title={Quantifying the creativity support of digital tools through the creativity support index},
  author={Cherry, Erin and Latulipe, Celine},
  journal={ACM Transactions on Computer-Human Interaction (TOCHI)},
  volume={21},
  number={4},
  pages={1--25},
  year={2014},
  publisher={ACM New York, NY, USA}
}

@article{flower1980,
 ISSN = {0010096X},
 URL = {http://www.jstor.org/stable/356630},
 author = {Linda Flower and John R. Hayes},
 journal = {College Composition and Communication},
 number = {1},
 pages = {21--32},
 publisher = {National Council of Teachers of English},
 title = {The Cognition of Discovery: Defining a Rhetorical Problem},
 urldate = {2025-10-03},
 volume = {31},
 year = {1980}
}

@article{flower,
 ISSN = {0010096X},
 URL = {http://www.jstor.org/stable/356600},
 author = {Linda Flower and John R. Hayes},
 journal = {College Composition and Communication},
 number = {4},
 pages = {365--387},
 publisher = {National Council of Teachers of English},
 title = {A Cognitive Process Theory of Writing},
 urldate = {2025-09-29},
 volume = {32},
 year = {1981}
}

@incollection{Kellogg,
  author    = {Ronald T. Kellogg},
  title     = {A model of working memory in writing},
  booktitle = {The Science of Writing: Theories, Methods, Individual Differences, and Applications},
  editor    = {C. Michael Levy and Sarah Ransdell},
  pages     = {57--71},
  publisher = {Lawrence Erlbaum Associates},
  address   = {Mahwah, NJ},
  year      = {1996}
}

@article{liu2023lost,
  title={Lost in the middle: How language models use long contexts},
  author={Liu, Nelson F and Lin, Kevin and Hewitt, John and Paranjape, Ashwin and Bevilacqua, Michele and Petroni, Fabio and Liang, Percy},
  journal={Preprint at arXiv},
  note={{https://doi.org/10.48550/arXiv.2307.03172}},
  year={2023}
}

@article{li2024long,
  title={Long-context llms struggle with long in-context learning},
  author={Li, Tianle and Zhang, Ge and Do, Quy Duc and Yue, Xiang and Chen, Wenhu},
  journal={Preprint at arXiv},
  note={{https://doi.org/10.48550/arXiv.2404.02060}},
  year={2024}
}

@inproceedings{li2023long,
  title={How Long Can Context Length of Open-Source LLMs truly Promise?},
  author={Li, Dacheng and Shao, Rulin and Xie, Anze and Sheng, Ying and Zheng, Lianmin and Gonzalez, Joseph and Stoica, Ion and Ma, Xuezhe and Zhang, Hao},
  booktitle={NeurIPS 2023 Workshop on Instruction Tuning and Instruction Following},
  year={2023}
}

@misc{chen2023walkingmemorymazecontext,
      title={Walking Down the Memory Maze: Beyond Context Limit through Interactive Reading}, 
      author={Howard Chen and Ramakanth Pasunuru and Jason Weston and Asli Celikyilmaz},
      year={2023},
      eprint={2310.05029},
      archivePrefix={arXiv},
      primaryClass={cs.CL},
      url={https://arxiv.org/abs/2310.05029}, 
}

@inproceedings{
sarthi2024raptor,
title={{RAPTOR}: Recursive Abstractive Processing for Tree-Organized Retrieval},
author={Parth Sarthi and Salman Abdullah and Aditi Tuli and Shubh Khanna and Anna Goldie and Christopher D Manning},
booktitle={The Twelfth International Conference on Learning Representations},
year={2024},
url={https://openreview.net/forum?id=GN921JHCRw}
}

@article{yao2024tau,
  title={$\tau$-bench: A Benchmark for Tool-Agent-User Interaction in Real-World Domains},
  author={Yao, Shunyu and Shinn, Noah and Razavi, Pedram and Narasimhan, Karthik},
  journal={Preprint at arXiv},
 note={{https://doi.org/10.48550/arXiv.2406.12045}},
  year={2024}
}

@article{barres2025tau,
  title={$\tau^2$-Bench: Evaluating Conversational Agents in a Dual-Control Environment},
  author={Barres, Victor and Dong, Honghua and Ray, Soham and Si, Xujie and Narasimhan, Karthik},
  journal={Preprint at arXiv},
  Note={{https://doi.org/10.48550/arXiv.2506.07982}},
  year={2025}
}

@article{wang2024appbench,
  title={AppBench: Planning of Multiple APIs from Various APPs for Complex User Instruction},
  author={Wang, Hongru and Wang, Rui and Xue, Boyang and Xia, Heming and Cao, Jingtao and Liu, Zeming and Pan, Jeff Z and Wong, Kam-Fai},
  journal={Preprint at arXiv},
note={{https://doi.org/10.48550/arXiv.2410.19743}},
  year={2024}
}

@misc{google_docs_gemini_collab,
  author       = {{Google Support}},
  title        = {Collaborate with Gemini in Google Docs (Workspace Labs)},
  howpublished = {\url{https://support.google.com/docs/answer/14206696?hl=en}},
  year         = {2025},
  note         = {Accessed: 2025-10-05}
}

@misc{google_docs_gemini_write,
  author       = {{Google Support}},
  title        = {Write with Gemini in Google Docs},
  howpublished = {\url{https://support.google.com/docs/answer/13951448?hl=en}},
  year         = {2025},
  note         = {Accessed: 2025-10-05}
}

@software{scrivener,
  author       = {Literature and Latte},
  title        = {Scrivener},
  year         = {2025},
  url          = {https://www.literatureandlatte.com/scrivener},
  note         = {Computer software}
}

@misc{overleaf,
  author = {John Hammersley and John Lees-Miller},
  title = {overleaf},
  year = {2025},
  publisher = {GitHub},
  journal = {GitHub repository},
  howpublished = {\url{https://github.com/overleaf/overleaf}}
}

@misc{openai2025introducinggpt5,
  author       = {OpenAI},
  title        = {Introducing GPT-5},
  year         = {2025},
  url          = {https://openai.com/index/introducing-gpt-5/},
  note         = {Accessed: 2025-10-02}
}

@misc{openai2025introducinggpt4-1,
  author       = {OpenAI},
  title        = {Introducing GPT-4.1 in the API},
  year         = {2025},
  url          = {https://openai.com/index/gpt-4-1/},
  note         = {Accessed: 2025-10-02}
}

@incollection{Olive2012,
  TITLE = {{Writing and working memory : A summary of theories and of findings}},
  AUTHOR = {Olive, Thierry},
  URL = {https://shs.hal.science/halshs-01367810},
  BOOKTITLE = {{Writing: A mosaic of new perspectives}},
  EDITOR = {Elena L Grigorenko and Elisa Mambrino and David D Preiss},
  PUBLISHER = {{Psychology Press}},
  YEAR = {2012},
  HAL_ID = {halshs-01367810},
  HAL_VERSION = {v1},
}

@article{guo2025deepseek,
  title={DeepSeek-R1 incentivizes reasoning in LLMs through reinforcement learning},
  author={Guo, Daya and Yang, Dejian and Zhang, Haowei and Song, Junxiao and Wang, Peiyi and Zhu, Qihao and Xu, Runxin and Zhang, Ruoyu and Ma, Shirong and Bi, Xiao and others},
  journal={Nature},
  volume={645},
  number={8081},
  pages={633--638},
  year={2025},
  publisher={Nature Publishing Group UK London}
}

@article{yang2025qwen3,
  title={Qwen3 technical report},
  author={Yang, An and Li, Anfeng and Yang, Baosong and Zhang, Beichen and Hui, Binyuan and Zheng, Bo and Yu, Bowen and Gao, Chang and Huang, Chengen and Lv, Chenxu and others},
  journal={arXiv preprint arXiv:2505.09388},
  year={2025}
}

@techreport{anthropic2025claude4,
  title        = {System Card: Claude Opus 4 \& Claude Sonnet 4},
  author       = {Anthropic},
  year         = {2025},
  month        = {May},
  url          = {https://www-cdn.anthropic.com/4263b940cabb546aa0e3283f35b686f4f3b2ff47.pdf},
}

@article{Edge2024FromLT,
  title={From Local to Global: A Graph RAG Approach to Query-Focused Summarization},
  author={Darren Edge and Ha Trinh and Newman Cheng and Joshua Bradley and Alex Chao and Apurva N. Mody and Steven Truitt and Jonathan Larson},
  journal={ArXiv},
  year={2024},
  volume={abs/2404.16130},
  url={https://api.semanticscholar.org/CorpusID:269363075}
}

@inproceedings{
rezazadeh2025from,
title={From Isolated Conversations to Hierarchical Schemas: Dynamic Tree Memory Representation for {LLM}s},
author={Alireza Rezazadeh and Zichao Li and Wei Wei and Yujia Bao},
booktitle={The Thirteenth International Conference on Learning Representations},
year={2025},
url={https://openreview.net/forum?id=moXtEmCleY}
}

@inproceedings{li-etal-2024-graphreader,
    title = "{G}raph{R}eader: Building Graph-based Agent to Enhance Long-Context Abilities of Large Language Models",
    author = "Li, Shilong  and
      He, Yancheng  and
      Guo, Hangyu  and
      Bu, Xingyuan  and
      Bai, Ge  and
      Liu, Jie  and
      Liu, Jiaheng  and
      Qu, Xingwei  and
      Li, Yangguang  and
      Ouyang, Wanli  and
      Su, Wenbo  and
      Zheng, Bo",
    editor = "Al-Onaizan, Yaser  and
      Bansal, Mohit  and
      Chen, Yun-Nung",
    booktitle = "Findings of the Association for Computational Linguistics: EMNLP 2024",
    month = nov,
    year = "2024",
    address = "Miami, Florida, USA",
    publisher = "Association for Computational Linguistics",
    url = "https://aclanthology.org/2024.findings-emnlp.746/",
    doi = "10.18653/v1/2024.findings-emnlp.746",
    pages = "12758--12786"
}

@article{comanici2025gemini,
  title={Gemini 2.5: Pushing the frontier with advanced reasoning, multimodality, long context, and next generation agentic capabilities},
  author={Comanici, Gheorghe and Bieber, Eric and Schaekermann, Mike and Pasupat, Ice and Sachdeva, Noveen and Dhillon, Inderjit and Blistein, Marcel and Ram, Ori and Zhang, Dan and Rosen, Evan and others},
  journal={arXiv preprint arXiv:2507.06261},
  year={2025}
}

@article{reza2025co,
  title={Co-writing with ai, on human terms: Aligning research with user demands across the writing process},
  author={Reza, Mohi and Thomas-Mitchell, Jeb and Dushniku, Peter and Laundry, Nathan and Williams, Joseph Jay and Kuzminykh, Anastasia},
  journal={arXiv preprint arXiv:2504.12488},
  year={2025}
}

@misc{google_docs,
  author       = {{Google}},
  title        = {Google Docs},
  howpublished = {\url{https://docs.google.com}},
  year         = {2025},
  note         = {Accessed: 2025-10-08}
}

@inproceedings{khurana2024and,
  title={Why and when llm-based assistants can go wrong: Investigating the effectiveness of prompt-based interactions for software help-seeking},
  author={Khurana, Anjali and Subramonyam, Hariharan and Chilana, Parmit K},
  booktitle={Proceedings of the 29th International Conference on Intelligent User Interfaces},
  pages={288--303},
  year={2024}
}

@article{stephen,
 ISSN = {0010096X},
 URL = {http://www.jstor.org/stable/356602},
 author = {Lester Faigley and Stephen Witte},
 journal = {College Composition and Communication},
 number = {4},
 pages = {400--414},
 publisher = {National Council of Teachers of English},
 title = {Analyzing Revision},
 urldate = {2025-10-06},
 volume = {32},
 year = {1981}
}

@article{nancy,
 ISSN = {0010096X},
 URL = {http://www.jstor.org/stable/356588},
 author = {Nancy Sommers},
 journal = {College Composition and Communication},
 number = {4},
 pages = {378--388},
 publisher = {National Council of Teachers of English},
 title = {Revision Strategies of Student Writers and Experienced Adult Writers},
 urldate = {2025-10-06},
 volume = {31},
 year = {1980}
}

@article{desmet2012,
title = {Write between the lines: Electronic outlining and the organization of text ideas},
journal = {Computers in Human Behavior},
volume = {28},
number = {6},
pages = {2107-2116},
year = {2012},
issn = {0747-5632},
doi = {https://doi.org/10.1016/j.chb.2012.06.015},
url = {https://www.sciencedirect.com/science/article/pii/S0747563212001653},
author = {M.J.R. {de Smet} and S. Brand-Gruwel and H. Broekkamp and P.A. Kirschner}
}

@article{desmet2011,
author = {Smet, Milou and Broekkamp, Hein and Brand-Gruwel, Saskia and Kirschner, Paul},
year = {2011},
month = {12},
pages = {557-574},
title = {Effects of electronic outlining on students' argumentative writing performance},
volume = {27},
journal = {J. Comp. Assisted Learning},
doi = {10.1111/j.1365-2729.2011.00418.x}
}

@article{limpo2018,
title = {Effects of planning strategies on writing dynamics and final texts},
journal = {Acta Psychologica},
volume = {188},
pages = {97-109},
year = {2018},
issn = {0001-6918},
doi = {https://doi.org/10.1016/j.actpsy.2018.06.001},
url = {https://www.sciencedirect.com/science/article/pii/S0001691818300672},
author = {Teresa Limpo and Rui A. Alves}
}

@article{klein2015,
author = {Klein, P.D. and Ehrhardt, J.S.},
year = {2015},
month = {01},
pages = {40-64},
title = {The effects of discussion and Persuasion writing goals on reasoning, cognitive load, and learning},
volume = {61},
journal = {Alberta Journal of Educational Research}
}

@article{fritz2025understanding,
  title={Understanding authorship in Artificial Intelligence-assisted works},
  author={Fritz, Johannes},
  journal={Journal of Intellectual Property Law and Practice},
  pages={jpae119},
  year={2025},
  publisher={Oxford University Press UK}
}

@article{mittelstadt2021interpretability,
  title={Interpretability and transparency in artificial intelligence},
  author={Mittelstadt, Brent},
  journal={The Oxford handbook of digital ethics},
  pages={378--410},
  year={2021},
  publisher={Oxford Academic}
}

@article{ai2023artificial,
  title={Artificial intelligence risk management framework (AI RMF 1.0)},
  author={AI, NIST},
  journal={URL: https://nvlpubs. nist. gov/nistpubs/ai/nist. ai},
  pages={100--1},
  year={2023}
}

@article{turpin2023language,
  title={Language models don't always say what they think: Unfaithful explanations in chain-of-thought prompting},
  author={Turpin, Miles and Michael, Julian and Perez, Ethan and Bowman, Samuel},
  journal={Advances in Neural Information Processing Systems},
  volume={36},
  pages={74952--74965},
  year={2023}
}

@article{siddiqui2025draftmarks,
  title={DraftMarks: Enhancing Transparency in Human-AI Co-Writing Through Interactive Skeuomorphic Process Traces},
  author={Siddiqui, Momin N and Nasseri, Nikki and Coscia, Adam and Pea, Roy and Subramonyam, Hari},
  journal={arXiv preprint arXiv:2509.23505},
  year={2025}
}

@article{ennab2022designing,
  title={Designing an interpretability-based model to explain the artificial intelligence algorithms in healthcare},
  author={Ennab, Mohammad and Mcheick, Hamid},
  journal={Diagnostics},
  volume={12},
  number={7},
  pages={1557},
  year={2022},
  publisher={MDPI}
}

@article{anakok2025interpretability,
  title={Interpretability of Graph Neural Networks to Assess Effects of Global Change Drivers on Ecological Networks},
  author={Anakok, Emre and Barbillon, Pierre and Fontaine, Colin and Thebault, Elisa},
  journal={arXiv preprint arXiv:2503.15107},
  year={2025}
}

@article{fan2021interpretability,
  title={On interpretability of artificial neural networks: A survey},
  author={Fan, Feng-Lei and Xiong, Jinjun and Li, Mengzhou and Wang, Ge},
  journal={IEEE Transactions on Radiation and Plasma Medical Sciences},
  volume={5},
  number={6},
  pages={741--760},
  year={2021},
  publisher={IEEE}
}

@inproceedings{lee2020itoc,
  title={iTOC: Enabling efficient non-visual interaction with long web documents},
  author={Lee, Hae-Na and Uddin, Sami and Ashok, Vikas},
  booktitle={2020 IEEE International Conference on Systems, Man, and Cybernetics (SMC)},
  pages={3799--3806},
  year={2020},
  organization={IEEE}
}

@article{gentner1983structure,
  title={Structure-mapping: A theoretical framework for analogy},
  author={Gentner, Dedre},
  journal={Cognitive science},
  volume={7},
  number={2},
  pages={155--170},
  year={1983},
  publisher={Elsevier}
}

@article{cao2025automating,
  title={Automating quantum computing laboratory experiments with an agent-based AI framework},
  author={Cao, Shuxiang and Zhang, Zijian and Alghadeer, Mohammed and Fasciati, Simone D and Piscitelli, Michele and Bakr, Mustafa and Leek, Peter and Aspuru-Guzik, Al{\'a}n},
  journal={Patterns},
  year={2025},
  publisher={Elsevier}
}

@article{2022creative,
  title={Creative Writing with an AI-Powered Writing Assistant: Perspectives from Professional Writers},
  author={Daphne Ippolito and Ann Yuan and Andy Coenen and Sehmon Burnam},
  journal={ArXiv},
  year={2022},
  volume={abs/2211.05030},
  url={https://api.semanticscholar.org/CorpusID:253420678}
}

@misc{bano2023investigating,
      title={Investigating Responsible AI for Scientific Research: An Empirical Study}, 
      author={Muneera Bano and Didar Zowghi and Pip Shea and Georgina Ibarra},
      year={2023},
      eprint={2312.09561},
      archivePrefix={arXiv},
      primaryClass={cs.AI},
      url={https://arxiv.org/abs/2312.09561}, 
}

@article{Resnik2024,
  title = {The ethics of using artificial intelligence in scientific research: new guidance needed for a new tool},
  volume = {5},
  ISSN = {2730-5961},
  url = {http://dx.doi.org/10.1007/s43681-024-00493-8},
  DOI = {10.1007/s43681-024-00493-8},
  number = {2},
  journal = {AI and Ethics},
  publisher = {Springer Science and Business Media LLC},
  author = {Resnik,  David B. and Hosseini,  Mohammad},
  year = {2024},
  month = may,
  pages = {1499–1521}
}

@article{khalifa2024,
title = {Using artificial intelligence in academic writing and research: An essential productivity tool},
journal = {Computer Methods and Programs in Biomedicine Update},
volume = {5},
pages = {100145},
year = {2024},
issn = {2666-9900},
doi = {https://doi.org/10.1016/j.cmpbup.2024.100145},
url = {https://www.sciencedirect.com/science/article/pii/S2666990024000120},
author = {Mohamed Khalifa and Mona Albadawy}
}

@misc{ramachandram2025,
      title={Transparent AI: The Case for Interpretability and Explainability}, 
      author={Dhanesh Ramachandram and Himanshu Joshi and Judy Zhu and Dhari Gandhi and Lucas Hartman and Ananya Raval},
      year={2025},
      eprint={2507.23535},
      archivePrefix={arXiv},
      primaryClass={cs.LG},
      url={https://arxiv.org/abs/2507.23535}, 
}

\appendix

\newtcolorbox{demobox}[2]{breakable,colback=#2!5!white,colframe=#2!75!black,fonttitle=\bfseries,title=#1}
\lstdefinestyle{base}{
  breaklines=true,
  basicstyle=\fontsize{7}{7}\selectfont\ttfamily
}
\definecolor{teal}{RGB}{50, 220, 220} 

\newtcolorbox{conversationbox1}[1][]{breakable,colframe=teal!5!black,
  colback=teal!75!white, width=\textwidth, fonttitle=\bfseries, title=#1}

\section{Participants of studies}

\label{app:participants}

The following tables shows the information of the participants in the formative (\autoref{tab:formative_study}), comparative (\autoref{tab:comparative_study}) and deployment study (\autoref{tab:deployment_study}) in this work. We note there is no overlap among the participants of the studies.

\begin{table}[h!]
\centering
\begin{tabular}{lll}
\toprule
\textbf{ID} & \textbf{Position} & \textbf{Field} \\
\midrule
F1 & PhD student & Physics \\
F2 & Postdoctoral researcher & Chemistry \\
F3 & PhD student & Bioinformatics \\
F4 & PhD student & Physics \\
F5 & Postdoctoral researcher & Computer Science \\
F6 & PhD student & Architecture \\
\bottomrule
\end{tabular}
\caption{Participants of the formative study, with their academic positions and fields of study.}
\label{tab:formative_study}
\end{table}

\begin{table}[h!]
\centering
\begin{tabular}{lll}
\toprule
\textbf{ID} & \textbf{Position} & \textbf{Field} \\
\midrule
P1  & Master’s student     & Computer Science \\
P2  & PhD student          & Computer Science \\
P3  & Undergraduate        & Computer Science \\
P4  & Undergraduate        & Computer Science \\
P5  & PhD student          & Chemical Engineering \\
P6  & Undergraduate        & Computer Science \\
P7  & Undergraduate        & East Asia Study \\
P8  & JD student           & Law \\
P9  & Master’s student     & Computer Science \\
P10 & PhD student          & Computer Science \\
P11 & Master’s student     & Computer Science \\
P12 & Undergraduate        & Computer Science \\
\bottomrule
\end{tabular}
\caption{Participants of the comparative study, with their academic positions and fields of study.}
\label{tab:comparative_study}
\end{table}

\begin{table}[h!]
\centering
\begin{tabular}{lll}
\toprule
\textbf{ID} & \textbf{Position} & \textbf{Field} \\
\midrule
L1 & PhD student & Computer Science \\
L2 & Postdoctoral researcher & Chemistry \\
L3 & Postdoctoral researcher & Chemistry \\
L4 & Postdoctoral researcher & Computer Science \\
L5 & Postdoctoral researcher & Chemistry \\
L6 & PhD student & Chemistry \\
L7 & Postdoctoral researcher & Chemistry \\
L8 & PhD student & Computer Science \\
\bottomrule
\end{tabular}
\caption{Participants of the field deployment study, with their academic positions and fields of study.}
\label{tab:deployment_study}
\end{table}

\section{Questions in the formative study}
\label{app:formative_question}

\section*{Interview questions}

The participants of the formative study is required to express their ideas on the following interview questions.

\paragraph{Background and Existing Practices}
\begin{itemize}
    \item Describe your typical process when working on a long document (e.g., a grant proposal).
    \item What do you find most challenging about writing or organizing long documents?
    \item What features do you wish writing tools provided to better support long or complex documents?
\end{itemize}

\paragraph{Perceptions of AI in Writing}
\begin{itemize}
    \item Have you used AI tools for writing tasks (e.g., drafting, summarizing, rewriting)?
    \item How did those AI tools impact your workflow?
    \item What would responsible AI support in collaborative writing look like to you?
\end{itemize}

\section*{Likert-scale questions}

After the interview questions, the participants are also required to answer the following Likert-scale questions (5 points).

\paragraph{Background and Existing Practices}

\begin{itemize}
    \item Writing long documents such as grants or papers is often frustrating.
    \item I often lose track of the overall structure when working on long documents.
    \item Existing tools (e.g., Google Docs, Overleaf) support organizing complex documents well.
    \item Navigating large documents with many sections and collaborators is difficult.
    \item Commented-out or outdated content makes it harder to stay focused or navigate.
\end{itemize}

\paragraph{Perception of AI in Writing}
\begin{itemize}
    \item I am comfortable using AI tools to assist with my writing.
    \item I would find it helpful if AI could suggest how to structure my content at a high level.
    \item I would trust AI to draft parts of my document based on my notes or instructions.
    \item AI-generated summaries of sections would help me stay organized and write better.
    \item I would feel more comfortable using AI if I could control which parts of the document it accesses.
\end{itemize}

\section{Post-task questions in study 1}
\label{app:study1-questions}
After the article modification task, the user is asked to answer the following Likert-scale questions (7 points).

\begin{itemize}
    \item I could easily obtain a clear overview of the document’s structure.
    \item Expanding outline points into paragraphs was easy.
    \item Breaking a long section into well-structured subsections was easy.
    \item Creating a summary from detailed content was easy.
    \item Propagating a key conceptual change consistently across a section was easy.
    \item The AI features were helpful and effective in assisting my writing tasks.
    \item I felt in full control over the edits and final content of the document.
\end{itemize}

After the creative writing task, the user is asked to answer the following Likert-scale questions (7 points).
\begin{itemize}
    \item The tool was effective for exploring and developing my initial ideas for the essay.
    \item The tool made it easy to build and refine the essay’s structure as my ideas evolved.
    \item It was straightforward to expand outline points into detailed paragraphs.
    \item The AI features were helpful and effective in assisting my writing tasks.
    \item I felt more productive and efficient in writing my essay using this tool.
\end{itemize}

\section{Experiment procedure of study 1}
\label{app:study1}

\begin{figure}
    \centering
    \includegraphics[width=\linewidth]{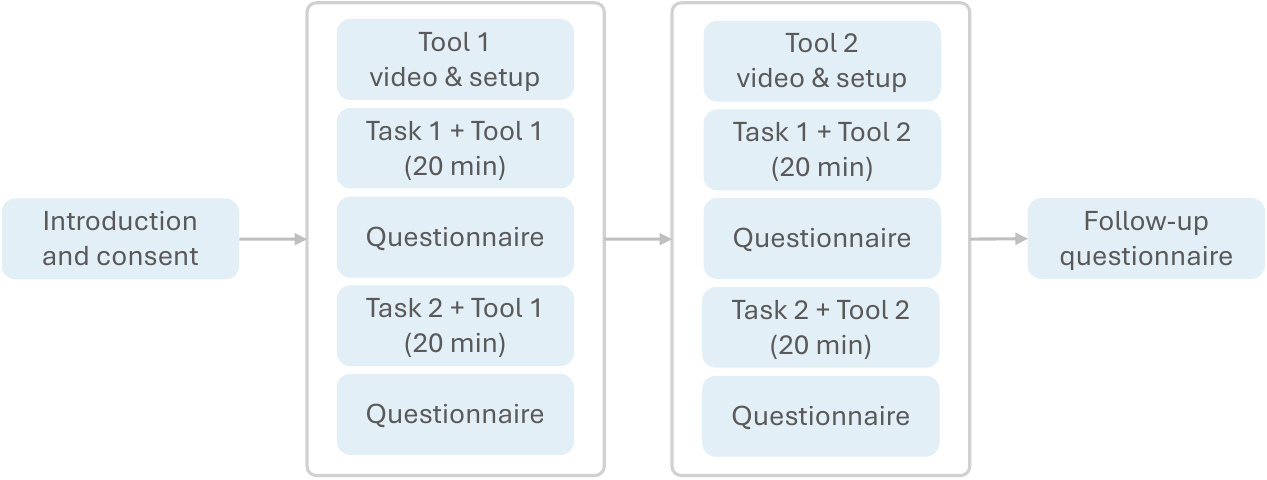}
    \caption{Overview of the experimental procedure of study 1.}
    \label{fig:experiment-procedure}
\end{figure}

Each user study session is split into two blocks. Each block contains two writing tasks to be completed by the user with either TreeWriter or the baseline editor. The order of tasks is randomized into four groups. Each user is randomly assigned a group before the start of the session in an effort to mitigate the impact of task ordering on experimentation results.

\begin{table}[h!]
\centering
\begin{tabular}{l c c c c}
\toprule
\textbf{Group} & \textbf{Activity 1} & \textbf{Activity 2} & \textbf{Activity 3} & \textbf{Activity 4} \\
\midrule
1 & TreeWriter + Article 1 & TreeWriter + Topic 1 & Baseline + Article 2 & Baseline + Topic 2 \\
2 & TreeWriter + Article 2 & TreeWriter + Topic 2 & Baseline + Article 1 & Baseline + Topic 1 \\
3 & Baseline + Article 1   & Baseline + Topic 1   & TreeWriter + Article 2 & TreeWriter + Topic 2 \\
4 & Baseline + Article 2   & Baseline + Topic 2   & TreeWriter + Article 1 & TreeWriter + Topic 1 \\
\bottomrule
\end{tabular}
\caption{The order of activity for each experimental group. }
\end{table}

The order of the tool block was counterbalanced between participants. Within each tool block, Task 1 was always conducted before Task 2 to minimize cognitive overhead from switching between editing and generative modes. To mitigate learning and carryover, we counterbalanced content: participants edited Article 1 (A1) in one tool block and Article 2 (A2) in the other, and they responded to different but matched prompts for the essays across the two blocks.

\section{Tasks for article modification}
\label{appendix:TasksForArticleModification}

The modification tasks use in the Task 1 are listed in \autoref{tab:task1}. Users were assigned either Article 1 or Article 2 in Task 1 of each tool block, depending on their group. Both articles were created by the authors starting with bullet points, which were then expanded into \prose using LLMs. In TreeWriter, the bullet points were retained within the document, while the generated \prose was exported via the export blocks. The Google Docs versions of the articles were produced by copying the linearized document in the linear view from TreeWriter.

\begin{table*}[h!]
\centering
\renewcommand{\arraystretch}{1.4}
\small
\begin{tabularx}{\textwidth}{>{\raggedright\arraybackslash}p{3cm} X X}
\toprule
\textbf{Design Goal} & \textbf{Article 1: AI’s Dual Role in Health Care} & \textbf{Article 2: AI’s Dual Role in Climate Change} \\
\midrule

\textbf{Content Generation and Elaboration} & 
\textbf{Location:} "AI Assistance for Health Literacy of Patients" \newline
\textbf{Task:} It currently provides only an outline. Expand this subsection into a full paragraph that explains how AI can support patients with different literacy levels, giving at least one concrete example of a feature based on the outline. &
\textbf{Location:} "Waste Reduction" \newline
\textbf{Task:} This section is concise and focused only on resource minimization. Expand it into a full paragraph that also gives one concrete example of how predictive scheduling can reduce waste in real industrial operations. \\
\midrule

\textbf{Content Adding} &
\textbf{Location:} Unknown \newline
\textbf{Task:} Integrate the following nuance into the paragraph while maintaining coherence: "Recent studies show that while AI improves early cancer detection, it also produces false positives that may lead to unnecessary treatments." &
\textbf{Location:} Unknown \newline
\textbf{Task:} Integrate the following nuance smoothly into the existing text: "Recent studies show that even highly accurate models can fail to predict compound disasters, such as simultaneous heatwaves and droughts, which are becoming more frequent under climate change." \\
\midrule

\textbf{Abstraction and Summarization} &
\textbf{Location:} "AI-Driven Patient Assistance and Access" \newline
\textbf{Task:} Create a new introductory paragraph that summarizes the key arguments across its subsections. Place the paragraph right after the section title. &
\textbf{Location:} "Enhancing Disaster Response" \newline
\textbf{Task:} Create a new introductory paragraph that synthesizes the key contributions across its subsections. Place the paragraph right after the section title. \\
\midrule

\textbf{Structural Reorganization} &
\textbf{Location:} "Automating Administrative Tasks in Healthcare" \newline
\textbf{Task:} The section is long and dense. Break it into three subsections with descriptive titles. Ensure each subsection contains the appropriate original content. &
\textbf{Location:} "Accelerating Battery Innovation and Green Energy Integration" \newline
\textbf{Task:} The section is long and covers multiple themes. Reorganize it into three subsections with descriptive titles and place the original content under each accordingly. \\
\midrule

\textbf{Conceptual Revision} &
\textbf{Location:} "Challenges and Risks of AI in Health Care" \newline
\textbf{Task:} Reframe the concept of “Over-dependent on AI” into “Abuse AI.” Replace all occurrences consistently, and update the section’s summary sentences to reflect this new framing. &
\textbf{Location:} "The Energy Cost of AI" \newline
\textbf{Task:} Reframe the concept of “energy consumption” into “carbon intensity.” Replace all mentions consistently, and update related summary statements in this section to reflect the new framing. \\
\midrule

\textbf{Coherence and Redundancy Cleanup} &
\textbf{Location:} "Balancing Benefits and Caution with AI" \newline
\textbf{Task:} Review paragraphs that overlap in describing “human expertise” and “critical thinking.” Use the AI system’s tools to harmonize tone, remove repeated statements, and ensure a smooth, coherent narrative flow. &
\textbf{Location:} "Balancing Act: Efficiency and Impact" \newline
\textbf{Task:} Review this section for repeated points about efficiency and sustainability. Use AI tools to harmonize tone, remove redundancies, and ensure the narrative flows smoothly. \\
\bottomrule
\end{tabularx}
\caption{Task 1 specification: Participants completed 6 modification tasks on the article designed to reflect realistic editing scenarios and to evaluate TreeWriter’s support for navigation, content generation, structural manipulation, conceptual editing and coherence management in line with DG1–DG2.}
\label{tab:task1}
\end{table*}

\clearpage
\section{Topics and prompts for creative writing}
\label{appendix:PromptsForCreativeWriting}

The following topics are provided to the user for the creative writing tasks.

\begin{itemize}
    \item[Topic 1:] The Role of Artificial Intelligence on Teaching in Universities
    \item[Topic 2:] The Impact of Artificial Intelligence on Students' Mental Health
\end{itemize}

The following prompt is also provided to the participant.

\begin{quote}
Write an opinion piece on the topic. Your essay should be well-structured and present a balanced view. It must include the following components:
\begin{itemize}
    \item An \textbf{introduction} that frames the issue and clearly states your main argument.
    \item A section discussing the \textbf{positive impacts}.
    \item A section analyzing the \textbf{negative impacts and potential risks}.
    \item A \textbf{conclusion} that synthesizes your points and proposes concrete directions for responsible adoption.
\end{itemize}
\end{quote}
\newpage
\section{Result of Study 1}
\label{app:study1-result}

\begin{figure}[htbp]
  \includegraphics[width=1\linewidth]{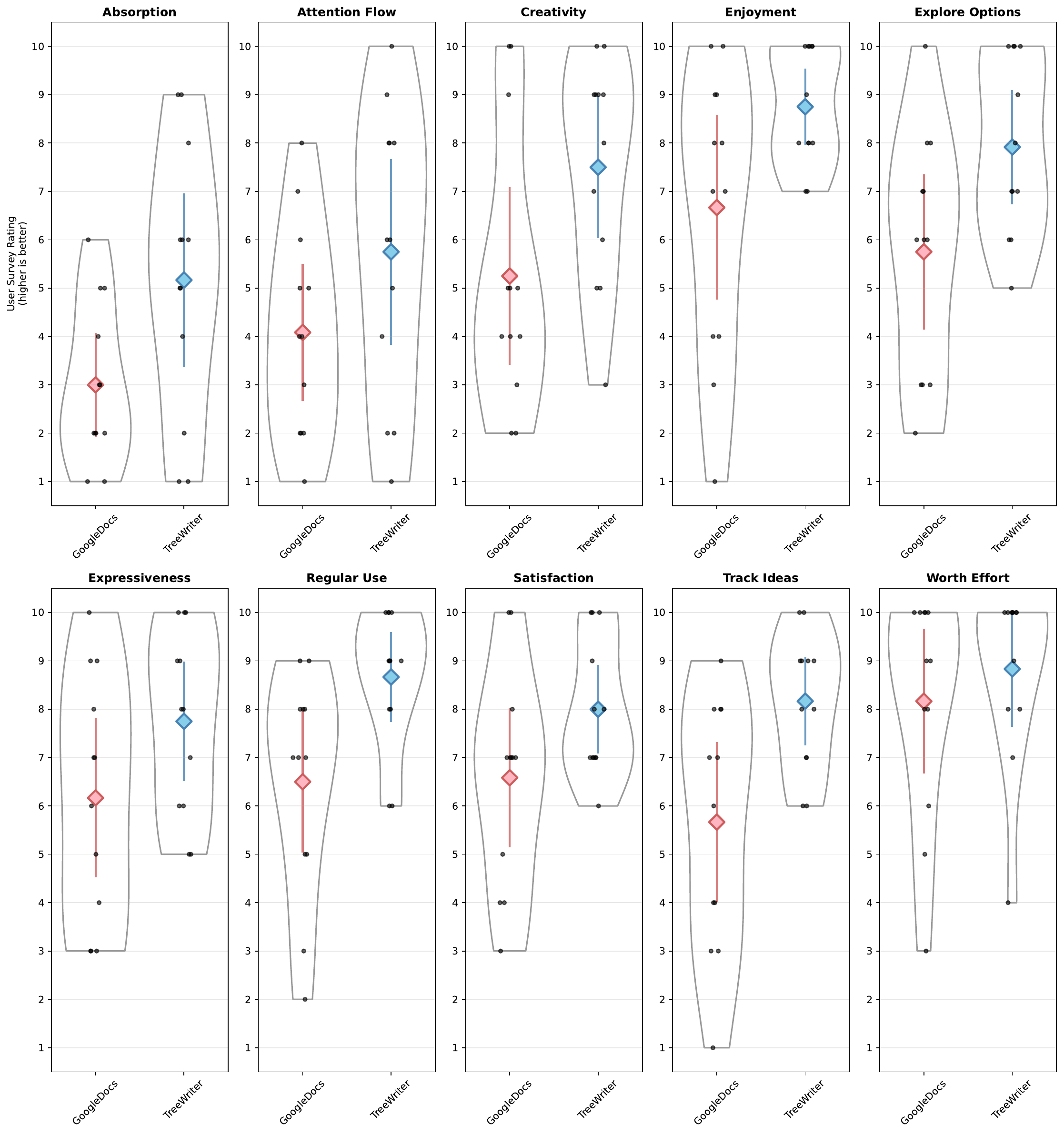}
  \caption{Violin plots comparing Creative Support Index (CSI) ratings between GoogleDocs and TreeWriter. The CSI is a standardized questionnaire that evaluates perceived
  creativity support across five dimensions: Exploration, Expressiveness, Immersion, Enjoyment, and Results Worth Effort. Participants completed the CSI after finishing the
   creative writing task. Each subplot shows the distribution of ratings (1-10 scale), with individual data points (black dots), mean values (colored diamonds), and 95\%
  confidence intervals (vertical lines).}
  \label{fig:csi_violin}
\end{figure}
\begin{figure}[htbp]
  \includegraphics[width=1\linewidth]{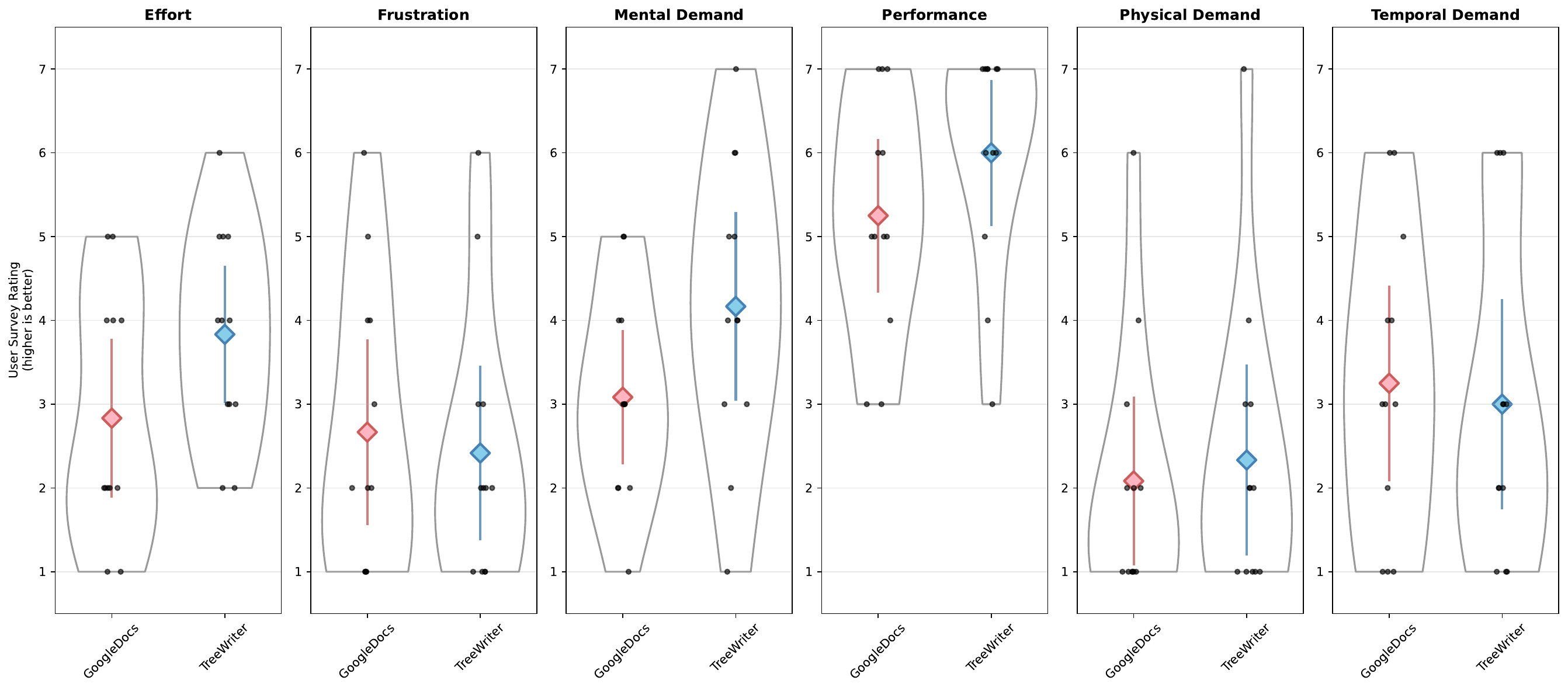}
  \caption{Violin plots comparing NASA Task Load Index (NASA-TLX) ratings between Google Docs and TreeWriter. The NASA-TLX is a standardized subjective workload assessment
  tool that measures six dimensions: Mental Demand, Physical Demand, Temporal Demand, Performance, Effort, and Frustration. Participants completed the NASA-TLX form after
  finishing Task 2, but the question asks about the participants' feelings about the tool rather than a single task. Each subplot shows the distribution of ratings (1-7 scale), with individual data points (black dots), mean
  values (colored diamonds), and 95\% confidence intervals (vertical lines). Lower scores indicate better performance except for the Performance dimension.
}
  \label{fig:nasatlx}
\end{figure}
 
In Study~1, TreeWriter consistently achieved higher scores in several key components, as indicated by Wilcoxon signed-rank tests. To address multiple comparisons, we controlled the false discovery rate \emph{within each measurement category} using the Benjamini--Hochberg procedure (FDR = 0.05; with $m$ denoting the number of tests in each category: CSI: $m=10$, NASA-TLX: $m=6$, Article modification post-task: $m=7$, Creative writing post-task: $m=5$). After within-category FDR correction, TreeWriter showed \textbf{statistically significant} advantages in \textit{Explore Ideas} ($q=0.01$), \textit{Creativity} ($q=0.033$), and \textit{Track Ideas} ($q=0.033$). 

While TreeWriter exhibited higher median scores in most categories, differences in some user-experience metrics, such as \textit{Frustration} and \textit{Mental Demand} from the NASA--TLX questionnaire, were less pronounced (see \autoref{fig:nasatlx}). Overall, these results suggest that users found TreeWriter to be a more effective and supportive tool, particularly for creative ideation and document structuring.

\newpage

\section{Questions in Study 2 and results}

\label{app:study2}

\begin{figure}
    \centering
    \includegraphics[width=1\linewidth]{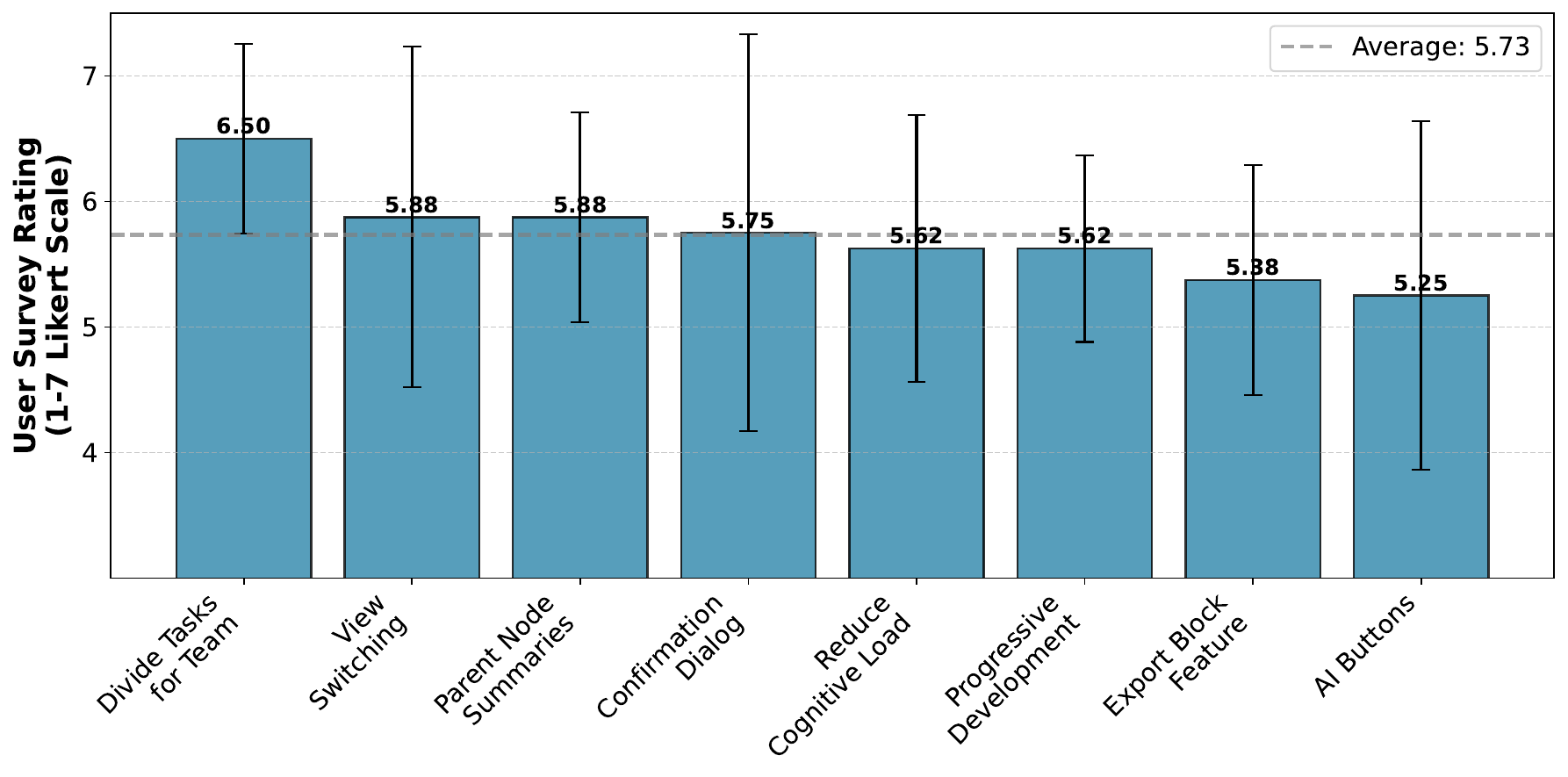}

    \caption{Average Likert-scale ratings (1–7) of key features/properties of TreeWriter from the field deployment study. Overall, participants rated them positively (M = 5.73). The team that received the highest score (M = 6.50) for 'Divide Tasks for Team' suggests strong support for features that facilitate collaboration. View Switching and Parent Node Summaries also scored above average (M = 5.88), indicating that users valued flexible navigation and hierarchical summarization. The lowest-rated feature was AI editing Buttons (M = 5.25), suggesting comparatively less perceived benefit from AI automation in this context. Error bars represent standard deviations.    \label{fig:deployment}}
\end{figure}

After the field deployment, the participants are asked to answer the following Likert-scale (1-7) questions, whose result can be found in \autoref{fig:deployment}. The participants also answered interview questions on the main advantages and disadvantages of using TreeWriter compared to Word or Google Docs, how the hierarchical and linear views affected their writing, which tasks TreeWriter suits best or worst, desired AI features, and any additional feedback.

\begin{enumerate}
    \item The tree structure helps me progressively develop my ideas from high-level concepts into detailed paragraphs.
    \item Using parent nodes to hold summaries or high-level notes helps me maintain a clear overview of my entire document.
    \item TreeWriter effectively reduces the cognitive load of managing a long, complex document.
    \item Switching between the Hierarchical Editor View and the Linear Preview View is a valuable part of my writing workflow.
    \item The AI editing buttons (generate paragraph/generate outlines/split paragraph) save my time in converting raw research ideas/results into paragraphs for paper.
    \item The confirmation dialog helps me review AI changes and gain better control in what is generated.
    \item I find the export block feature to be a clear and useful way to separate my private notes/outlines from the final, publishable text.
    \item The tree structure makes it easy to divide writing tasks and assign responsibility for different sections to team members.
\end{enumerate}

\section{Prompts for AI functionality}

\subsection{Writing assistant}
\label{app:assistant_prompt}
\begin{demobox}{Prompt for writing assistant}{teal}
\begin{lstlisting}[style=base]
You are a professional writing assistant AI agent that helps user write on a tree structure documents, where each node is a unit of content.

Your context is the following:
<context>
Currently, the user select node (ID: ${nodeM.id}):
<currentContent>
${originalContent}
</currentContent>

${parentContent}

${markedNodeContent}

${childrenInfo}

${siblingsInfo}
</context>

You are here to help the user with writing tasks including:
- Improving and refining existing content
- Providing writing suggestions and feedback
- Generating new content based on user requests
- Answering questions about the current content

Terminology:
- A "node" refers to a single unit of content in the tree structure.
- A "level" refers to all the siblings of the node and the node itself.
- A "section" refers to a node and all its descendants.

Logic of tree structure:
- The parent node should be a summary of all its children nodes by default.
- The children nodes should be more detailed and specific than the parent node.
- Bullet points are preferred for any nodes unless the user specifies otherwise.
- The <div> with class "export" is a special container that contains the content generated from other content in that node. It should be considered as a part of the node.

Task instructions:
- "Matching children": you should check whether the current children nodes reflects the content in the current node. If not, you should suggest modifying the children nodes to match the content in the current node.
- "Split into subsections": you should create new children nodes with proper titles to cover all the information of the current node.
- "Write paragraph": if the target node has bullet points, you should write a paragraph based on them and add an <div class="export">...</div> to contain the paragraph in the end of the node.
- "Revise section": you should consider revising the content and title of the current node, as well as its children nodes if exists.

You must 
- You must use tool loadNodeContent to get the content of a node first before writing about them.
- If the user asks for writing something, by default, it means that you need to call suggestModify tool to write the content for the current node.
- Don't drop the links in the content. Put every <a></a> link in a proper place with proper content.
- Don't expand an abbreviation by yourself.
- If the user specifies another node, you can also call suggestModify tool to write for that node.
- You don't need to mention what new version you created in your text response, as the user will see the new version directly.

Keep in mind:
- The user is viewing the sibling nodes. If they mention nodes in plural, they might want you to consider the sibling nodes.
- Always use tools to suggest changes. Never just write your suggestions in the text response.
- You can use the search tool to find more nodes if the current context is not enough.

You can create new versions for any nodes you know id - the current node, parent node, marked nodes, children nodes, or sibling nodes. This allows you to suggest improvements to related content beyond just the current node.

Respond naturally and conversationally. You can include regular text explanations along with any new content versions using the tool. Focus on being helpful and collaborative in your writing assistance.

\end{lstlisting}
\end{demobox}

\subsection{AI-powered editing buttons}
\label{app:ai_button_prompt}

\begin{demobox}{Prompt of ``split into subsections'' button}{teal}
\begin{lstlisting}[style=base]
You are a professional editor. Your task is to break a long content into multiple children nodes.
Generate a list of titles for new children nodes that completely cover the original content.

<original_content>
${parentContent}
</original_content>

Please distribute the parent content into multiple new children nodes, each with a title.
If there is any existing sectioning logic in the original content, please respect them. For example, if there are sublists or bold headings, use them as sectioning to create new children nodes.
The titles should be concise and descriptive.
You should add at most 5 new children nodes. Therefore, you should first analyze where the break the original content.
<output_format>
You should only return a JSON array of JSON objects with the following format.
There should be at most 5 objects in the array.
You should not put any HTML elements not appearing in the original content.
[
    {"title": <A concise and descriptive title>, "content": <HTML content for the new child node 1>},
    ...
]
</output_format>
\end{lstlisting}
\end{demobox}

\begin{demobox}{Prompt of ``generate paragraph'' button}{teal}
\begin{lstlisting}[style=base]
You are a professional writer. Your task is to write well-written paragraph(s) based on the raw content provided.

${nodeM.title() ? `
<node_title>
${nodeM.title()}
</node_title>
` : ''}

<raw_content>
${contentExceptExports}
</raw_content>

<current_paragraphs>
${existingExportContent}
</current_paragraphs>
${userPrompt ? `
<user_instructions>
${userPrompt}
</user_instructions>
` : ''}

Writing guidelines:
- You must write a paragraph based on the raw content.
- You should not expand an abbreviation unless it is expanded in the raw content.
- You should keep the <a></a> links in the raw content and put them in a proper place.

You must:
- If the current paragraph is not empty, don't make unnecessary changes. You are encouraged to keep the original contents as much as possible.
- If there is a mismatch between the current paragraphs and the raw content, make paragraphs align with the raw content.
- You should not drop any key information in the raw content.
- Be written in a clear, professional style
${userPrompt ? '- Follow any additional instructions provided above' : ''}

<output_format>
You should only return the HTML content of the paragraph without any additional text or formatting.
You should keep the links in the original content and put them in a proper place.
</output_format>
\end{lstlisting}
\end{demobox}

\begin{demobox}{Prompt of ``generate outlines from children'' button}{teal}
\begin{lstlisting}[style=base]
You are a professional editor. The given <original_content> is a summary of some children contents. 
Your task is to revise the <original_content> to make it serve as a better summary of the children contents.
<original_content>
${originalContent}
</original_content>
<children_contents>
${childrenContent}
</children_contents>
Please provide an updated version of the <original_content> in key points if <original_content> is not empty. You should not make unnecessary changes to the <original_content>.
<output_format>
You should only return the HTML content of the revised text without any additional text or formatting.
You are required to make a list of key points by <ul></ul>
You are encouraged to use <strong></strong> to highlight important information.
You can make sub-lists using <ul> within a <li>.
You should make no more than 5 key points. Each key point should be less than 30 words.
If there are any annotations in the original text, you should keep them as they are.
</output_format>
\end{lstlisting}
\end{demobox}

\end{document}